# Field-control, phase-transitions, and life's emergence

**Gargi Mitra-Delmotte**[1*] **and A.N. Mitra**[2*]

[1]39 Cite de l'Ocean, Montgaillard, St.Denis 97400, REUNION.
e.mail : gargijj@orange.fr

[2]Emeritius Professor, Department of Physics, Delhi University, INDIA; 244 Tagore Park, Delhi 110009, INDIA;
e.mail : ganmitra@nde.vsnl.net.in

***Correspondence:**
Gargi Mitra-Delmotte[1]
e.mail : gargijj@orange.fr

A.N. Mitra[2]
e.mail : ganmitra@nde.vsnl.net.in

Number of words: ~11511 (without Abstract, references, tables, and figure legends)
7 figures (plus two figures in supplementary information files).

**Abstract**

Instances of critical-like characteristics in living systems at each organizational level (bio-molecules to ecosystems) as well as the spontaneous emergence of computation (Langton), do suggest the relevance of self-organized criticality (SOC). But extrapolating complex bio-systems to life's origins, brings up a paradox: how could simple organics--lacking the 'soft matter' response properties of today's complex bio-molecules--have dissipated energy from primordial reactions (eventually reducing $CO_2$) in a controlled manner for their 'ordering'? Nevertheless, a causal link of life's macroscopic irreversible dynamics to the microscopic reversible laws of statistical mechanics is indicated via the 'functional-takeover' of a soft magnetic scaffold by organics (c.f. Cairns-Smith's "crystal-scaffold"). A field-controlled structure offers a mechanism for bootstrapping--bottom-up assembly with top-down control: its super-paramagnetic colloidal components obey reversible dynamics, but its dissipation of magnetic (H)-field energy for aggregation breaks time-reversal symmetry. The responsive adjustments of the controlled (host) mineral system to environmental changes would bring about mutual coupling between random organic sets supported by it; here the generation of long-range correlations within organic (guest) networks could include SOC-like mechanisms. And, such cooperative adjustments enable the *selection* of the *functional* configuration by altering the inorganic dipolar network's capacity to assist a spontaneous process. A non-equilibrium dynamics could now drive the kinetically-oriented system (trimming the phase-space via sterically-coupled organics) towards a series of phase-transitions with appropriate organic replacements "taking-over" its functions. Where available, experiments are cited in support of these speculations and for designing appropriate tests.

## 1 Introduction

The implications of minerals in life's emergence were first envisaged by Goldschmidt (1952) and Bernal (1949); these included concentration (adsorption) and catalysis, besides chirality of organics via association with crystal-surfaces. This motivated many works (see Arrhenius 2003; Carter 1978; Ferris 1999; Hazen and Sverjensky 2010; Jacoby 2002; Lahav 1999; Lambert 2008; Schoonen et al 2004; Seigel and Seigel 1981, and references therein), and inspired scenarios exploring the resemblance of ancient enzyme-clusters to mineral ones in metabolism-first approaches to life's origins (Cody et al 2000; McGlynn et al 2009; Russell and coworkers (Sect.6.3); Wachtershauser 1988). Hazen (2006) reviews the role of mineral surfaces for assistance at two stages of increasing complexity, viz. (1) emergence of biomolecules, and (2) emergence of macromolecular systems. These in turn cover three aspects: (i) possible enhanced self-assembly of lipids in the presence of minerals (Deamer and Pashley 1989; Luisi 1989; Hanczyc et al 2003; Chen et al 2004); (ii) polymerization of amino acids and nucleic acids, (Sowerby 1996; Uchihashi et al 1999; Lahav et al 1978; Ferris 1993; Liu and Orgel 1998; Orgel 1998) where Smith (1998) uses channels of zeolites as a packing constraint to help polymerization; and (iii) selective adsorption onto mineral surfaces, of organics (Churchill et al 2004; Carter 1978; Lowenstam and Weiner 1989). The latter include chiral molecules (Lahav 1999; Jacoby 2002; Hazen and Sholl 2003), although Hazen (2006) also mentions other mechanisms for chiral selection like determinate vs chance local processes. Apart from parity violation in beta-decay; he considers chiral-selective photolysis by circular-polarized synchrotron radiation from neutron stars (Bailey et al. 1998; Podlech 1999); magnetochiral photochemistry (Rikken and Raupach 2000); and at smaller scales the amplification of slight chiral excesses via Bose-Einstein condensation (Chela-Flores 1994), or chiral self-assembly of polymers (Bolli et al. 1997; Lippmann and Dix 1999; Saghatelian et al. 2001) or simply crystals (Eckert et al. 1993; Lahav and Leiserowitz 1999).

According to Hazen (2006), Cairns-Smith's (1968) theory is the most extreme form of mineral-based hypotheses positing that clay crystals were the precursors of today's replicators. As we see it, in this two-level scenario, the *hosting* inorganic layer or the crystal-organization—call it level-I (depicted as a white pin board in Figure 1a), -- offers top-down control and assistance for the bottom-up assembly of organic materials into complex patterns building up from *randomly* reacting/interacting entities in the '*guest*' layer—call it level-II (depicted with coloured beads, lower Figure 1a). In the latter, chemical reactions lead to building blocks, small polymers, proto-metabolic reactions, etc. while weak physical interactions (e.g. Hunding et al 2006) lead to small assemblies. Now, level-I's own crude *functional organization* acts as a selection/'trimming' mechanism for 'fishing out' constructs with superior function (information-propagation capacity) from the multitude of species forming at level-II. This leads to a gradual replacement of the inorganic organization by organic modules (coloured pattern, upper Figure 1a), whose recruitment by a functional system --aided by complementary interactions-- is crucial for their dynamic stability (see Sect.3); conceptually too, this *relates structure of the organic module to its function*. Also, level-II products favouring propagation of template-information (level-I) enable *feedback* between the levels. But, unlike hard crystals, a soft fractal organization seems a more natural origin for bio-complexity. To that end, a colloidal-gel scaffold (Sect.6) seems promising as a dynamically stable confining medium compatible with the key role of diffusion-controlled reactions in cellular biochemistry (Kopelman 1989; Konkoli 2009). A gradual 'takeover' by organic modules is also easier to visualize via a dynamic inorganic modular organization, e.g. soft colloids (Russell et al 1990), provided one can associate them with a crystal-like organization, towards a formal theory.

Now, unlike mineral-based bottom-up approaches adhering to the "metabolism-first" camp, the crystal-scaffold theory proposes a pre-existing template-organization, thus upholding the "genes-first" one. The former tells how *local* mineral-organic interactions can assist guest-level-II reactions, while the latter considers *global* aspects, i.e., bio-like functions linked to a cooperative organization of mineral-hosts. Indeed, these are complementary, and roughly correspond to the 2-tier organization of living systems: the control-network-level-I of complex biomolecules (proteins, nucleic acids, lipids, carbohydrates, etc) maps to the hosting functional mineral-organization, and the metabolic network-level-II maps to the (guest) organic reactions/interactions. In this federal-like anatomy of a living system, each tier/sub-system functions independently—albeit constrained by feedback-coupling (c.f. life's irreducible structure (Polyani 1968)). Now, the second correspondence-- between guest reactions/interactions and metabolic-network-level-II-- is easier to visualize than the first one (Sect.3). Indeed, while *macroscopic energy flow* in the metabolic reaction cycles can be mapped to that in similar organic attractors in abiogenesis, we still need a mapping -- albeit in terms of inorganic matter-- for the control-network (level-I) capable of *microscopic energy transactions*. This can be seen at the level of the *components* undergoing infinitesimal conformational changes to traverse a continuous energy landscape, or even at the global *system* level, where diverse closely spaced states in genotype-space are accessible via environmental fluctuations. Sure enough, open living systems can *harness fluctuations* -- at component (for work-cycles) and (evolving) system levels—unlike technological devices, sealing off external noise.

Now, mechanisms consuming free energy in the least time provide a natural basis for energy flows to select (pre-biotic/genetic) amongst dissipative structures (Annila and Salthe 2012). And selection could have started on "technologically simpler" (Cairns-Smith 2008) variations in energy channeling mechanisms (c.f. complex organic functional networks building up from scratch): Analogous to interdependent metabolism and replication, in abiogenesis, *a spontaneous process* provides a thermodynamic incentive for sustaining the continuity of an *environment-coupled cooperative network* assisting its occurrence (function). Complex replacements of the network could have arisen via different mechanisms, including self-organized criticality (SOC; Sect.2) among candidates despite its lack of a predictable framework. Note that otherwise in the origins of life its role seems limited if only guest-level-II processes are considered, since proto-metabolic reactions or weak interactions between organics dispersed in random mixtures alone cannot suffice for SOC to be effective. [As for relative orders of magnitude, bond energies involved in covalent bonds vs those for Van der Waals clusters bear the ratio: several eV vs a fraction of an eV (Kreuzer 2005), compared to thermal energy ($k_BT$) of

~ hundredth of an eV]. In the absence of an instructional principle, a random process of putting together simple organic building blocks (or mineral-particle-bound ones), into an intricate informational system would seem futile in view of the negligible probabilities at each step, for one wonders what interactive mechanisms are needed to ensure that random mixtures be *stable* enough to stay together to facilitate long-range correlations between them. Thus to reconcile the slow evolution of pure thermodynamic processes with the faster one of life's, a mechanism (e.g. trimming phase-space) is needed to break free from the constraints of thermodynamics, while paying obeisance to it. To that end, inspired by Cairns-Smith's pre-existing crystal-organization, we look to the signatures of *fields* on some collectively interacting entities at the inorganic-host-level-I that could have conferred on them the capacity to assist in the advanced stages of complexity, viz. emergence of replicators evolving via natural selection (Hazen 2006). In particular, the advantage of an external H-field-cum-magnetic nano-particles (MNPs), vis-à-vis say an electric field controlled system of particles, relates to the *diamagnetic properties of nano-sized organics*; thus anchoring the latter to mineral colloids responsive to an H-field can be used as an indirect means to exert control on them (there being associated dielectric properties with both mineral and organic colloids).

Among possible scenarios, one may consider its potential to give its responsive nano-scale materials 1) a dynamical basis of orientation in a liquid phase enabling formation of aggregates due to dipolar interactions (Taketomi 2011), leading to 2) a *response to the external by the generated internal field* in the interacting system (Huke and Lücke 2004; Sect.4), whose global evolution would also depend on the susceptibility of its materials to external factors (such as temperature). This quintessential analogue-information system (Palm and Korenivski 2009) seems plausible as a scaffold for the emergence of life as it has potential for cooperative interactions at two levels: a) a colloid component, whose spins (exchange-coupled in MNP lattice) constitute the particle's composite spin, and b) the dipolar interactions between the components themselves. Indeed, anisotropic dipolar interactions in fluids impinge on fundamentals, such as direction dependence, intrinsic long range nature and susceptibility to external forces (Wei and Patey 1992; Weis et al 1992; see Klapp 2005 plus references; Sect.4), and are biologically intriguing (Tavares et al 1999). In ferrofluids (single-domain MNP suspensions in carrier liquids), these can lead to correlations between neighbouring dipoles in growing fractal clusters, wherein to minimize dipolar energy, dipoles prefer to be parallel head-to-tail or antiparallel side-to-side (Pastor-Satorras and Rubí 1995; 1998). In zero-field (for particles with large magnetic moments) dipolar interactions can lead to isotropic fractal aggregates, qualifying them as SOC systems (no external driving). Thus the expected field-induced scaling behaviour was described as the response of this fractal equilibrium system at the critical point to the small external field conjugated to its order parameter (Botet et al 2001). Now, the change from zero-field with diffusion-limited aggregation to a field-driven one in moderate fields is expected to reduce the fractal dimension of the reversible structures (c.f. micro-particles, Domínguez-García and Rubio 2010). The intermediate regime suggests an access to statistical features like *scaling* on the one hand, as well as *controlled mobility* on the other, via field-control. These ingredients offer a confined biological-like (level-I) system with potential for feedback effects: its *susceptible* global-configuration—dictating function-- cannot be determined from properties of its components alone, and it can *influence* the orientations/dynamics of sterically-coupled organics at guest-level-II (Sect.5). Here, responsive adjustments to changing external influences (via size and magnetic moment of incoming MNPs, reactions or interactions at level-II, fluxes, etc) can affect the network's capacity to 'function' (say transport, Sect.4.6), thus providing a basis for selection of a configuration. The potential to collectively respond to external changes seems an important requirement for a hosting-scaffold (level-I) in view of the penetrating influence of the environment upon a living system whose internal state adjusts to changes in the former.

Further, the ability of far-from-equilibrium living systems to act as conduits of energy flow equips them with dynamic stability. The construction of their dynamical nano-components/subsystems calls for a scaffold-medium with reversible interactions enabling their interplay with external fluxes, thermal motion, etc. (c.f. convergence of energies, Fig.2 in Phillip and Quake (2006)) Such a *controlled* organization with reversible dynamics (Whitesides and Boncheva 2002) —as a starting-point for a cell-like organization-- seems inaccessible to a host medium with irreversible linkages such as rock pores or thermally linked inorganic gels (despite their importance for generating abiogenics, or compatibility with other magnetic/physical effects). Again, in contrast to organics randomly floating within aqueous spaces entrapped in liposomal sacs or rock pores, the suggested *flow-reactor-type scenario* enables *association only of entities actively coupled* with the field-controlled system, such as organics bound to mineral-particles, or those interacting with bound organics, etc. (Sect.3.3). To that end the microfluidic system by Park and Kim (2010) seems promising. Furthermore, Ranganath and Glorius (2011) draw attention to the advantages of using externally-controllable super-paramagnetic particles in a range of applications -- from quasi-homogeneous catalytic systems to data storage. Figure 1b depicts the idea that a field-controlled and dynamically stable inorganic modular organization (c.f. Cairns-Smith 1985) can i) support the gradual evolution of organic mixtures at guest-level-II, ii) be compatible with the simultaneous emergence of different kinds of organic networks/autocatalytic subsystems (c.f. Gánti's (2003) 3 sub-systems), iii) *simultaneously affect any coupled subsystems* and thus hasten their mutual cooperative interactions, thanks to the influence of the environment on its own *d.o.f.s,* e.g. via an SOC mechanism, and iv) by virtue of its capacity for some *primitive functions*, provide a *selection basis* towards its own 'takeover' by superiorly functioning organic networks. Note that this crucial role envisaged for an inorganic functional scaffold only concerns the initial stages of life's emergence, for providing a feedback circuit between levels I and II till both became organic-based.

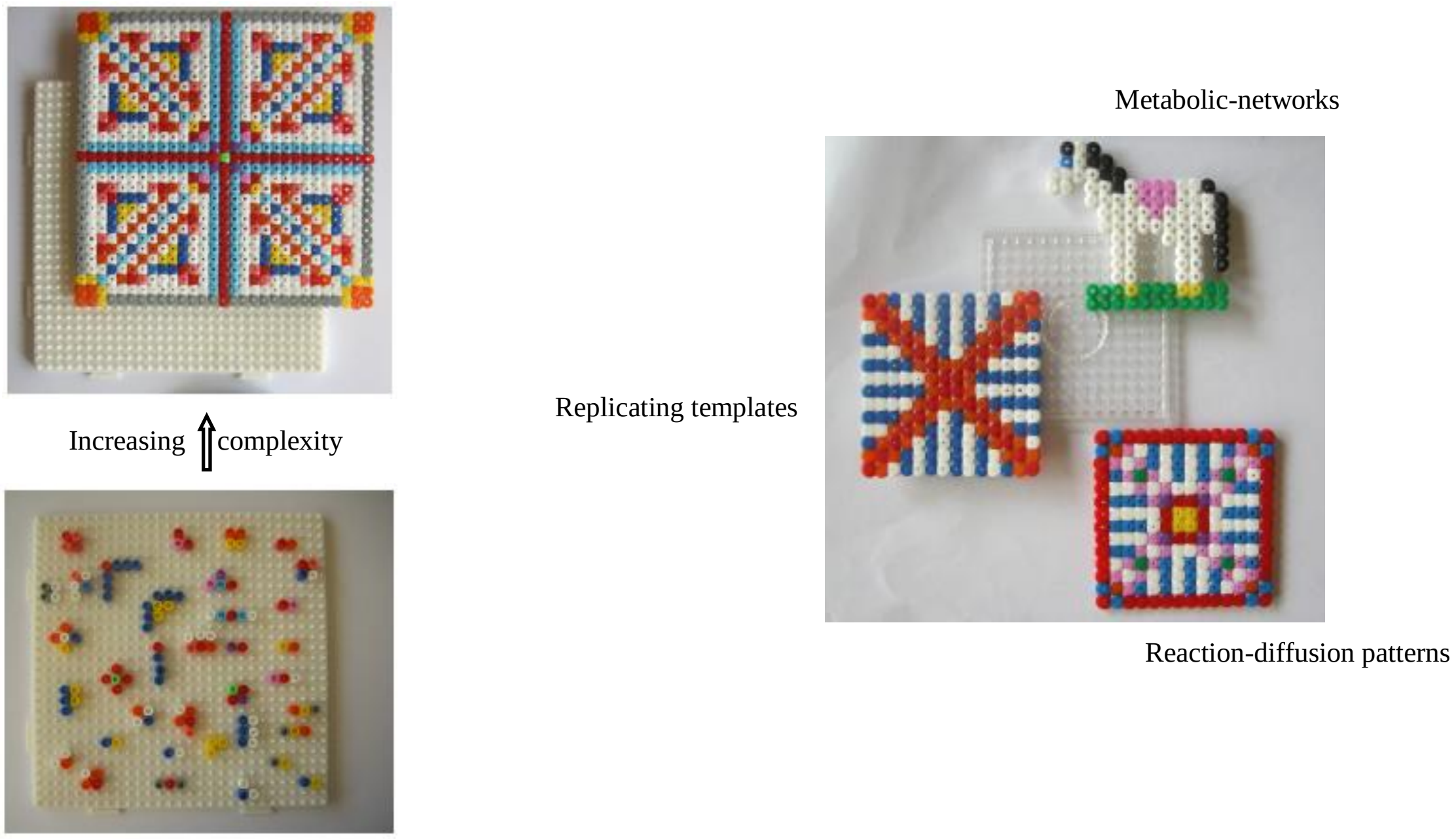

(a) (b)

**Figure 1**

To get an intuitive feel for the organizing power of a field, think of system-components as compasses detecting/responding to magnetic field lines, or iron filings showing the lines of force from a bar magnet. Similarly liquid-dispersed magnetic nano-particles form north-to-south chains, joining together end-to-end, while adjacent strings show a repelling property. In a similarly polarized ferrofluid (http://en.mobile.wikipedia.org/wiki/File:Ferrofluid_poles.jpg) this particle alignment effect is spread uniformly throughout the liquid medium and a sufficient field for overcoming gravity/surface tension can make spikes appear (e.g. see Peter Terren's website http://tesladownunder.com/Ferrofluid.htm). In fact, the remarkable similarity of magnetic/electric fields on MNP/thread suspensions, respectively, to the mitotic spindle, led Rosenberg et al (1965) to study the effect of fields on cell division and related applications. Also, the dimensions of a cell~ 10 -100 micrometer; protein ~5-50 nm; gene~2nm wide and 10 -100nm long (Pankhurst et al 2003), show that MNPs have the same length scale as biomolecules, thus making it possible to apply magnetic-field induced clustering and cell signaling using these tiny magnets as ligands (Mannix et al 2008; see also Chen 2008), and also enhance the potential of field-effects in origins-of-life research. The crucial role of fields in biology today underlying cooperative effects (see Ho 1997 and ref), also provides a natural motivation to look for *coherent* influences in the origins of life that could have caused *cooperative interactions*.

In this review, Sect.2 considers the implications of SOC in life's emergence, after a brief look at biological-systems and SOC. Sect.3 studies Cairns-Smith's "crystal-scaffold" organization using an LC-medium, and the potential of a soft-scaffold for assisting bottom-up approaches via kinetic aspects. Towards a 'boot-strapping' scenario, we briefly look at field-induced dynamical structures in Sect.4 to see how field-control can cause confinement of particles, influence their global configuration, and render them as carriers for transferring heat and electrons. Sect.5 studies how these controlled-systems could have caused cooperative transitions in organic-matter. Sect.6 briefly considers fractal structures and their implications for harnessing gradients, and studies the hydrothermal mound scenario with potential for forming such structures, before conclusions in Sect.7.

## 2 Living systems and SOC; implications of SOC in life's emergence

### 2.1 Living systems and SOC

Biological systems are self-organizing systems with a globally coherent pattern emerging spontaneously, thanks to the cooperative local interactions of its components. Important universal facets include: 1) *distributed control*, with all elements functioning as independent units in parallel, e.g. heterarchy in an ant colony (Dréo and Siarry 2004); 2) controlled work-cycles of nano-machine *components*; for example, motors require a slow input from a non-equilibrium source (homogeneous) plus rectified thermal fluctuations, thanks to the *asymmetric* nature of their surfaces appropriate for ratchet dynamics (Astumian and Derenyi 1998; Astumian and Hangii 2002); 3) controlled global dynamics of the *system* undergoing slow and adaptive alterations in response to environmental fluctuations; 4) chirality and polar asymmetry of building blocks for asymmetric dynamics; and 5) fractal (nested) nature of organization (Ho 1997), enabling components to locally operate close-to-equilibrium (see point 2) with optimal efficiency despite staying globally far-from-equilibrium. A similar fluctuation-driven formation of order from disorder is a familiar phenomenon in equilibrium systems undergoing phase transitions (see Box-I)--a typical form of spontaneous symmetry breaking. Note that potential energy is an integrated effect of interactions of specific arrangements (e.g. parallel/anti-parallel spins), signifying order, unlike fluctuations that characterize disorder. And spontaneous symmetry-breaking means that despite the system's equations of motion being symmetrical, the instability in the internal chemistry of its components, causes a loss of homogeneity/symmetry to the system's state (Anderson and Stein 1985). Transitory self-organized patterns are also seen in turbulent thermodynamic systems far-from-equilibrium, e.g. convection but they do not match those of robust living systems that exhibit stability and control at each point of their dynamics, despite dissipating energy and creating entropy to maintain their structure (Anderson and Stein 1985). Again, in vortices, typically macroscopic perturbations or higher-level structures do not modify the (internal) structure of the molecular components, unlike the bi-directional informational flows between different levels of bio-organization (Hartwell et al 1999). On the other hand, the fractal patterns in diffusion-limited aggregation (DLA) processes are somewhat reminiscent of *structural* complexities of their bio-counterparts, especially in the transporting role of diffusion (Witten and Sander 1981).

The analogy to slowly evolving living systems becomes clearer for certain slowly driven non-equilibrium systems that can "self-organize" into a robust stationary state with a scale-invariant macroscopic behaviour, owing to dissipative transport processes associated with a critical variable (Bak et al 1987; 1988). This phenomenon--dubbed as self-organized criticality (SOC) —shares some commonalities with the equilibrium concept of second order phase transition (see Box-I), usually associated with scale-invariance, maintained by fine-tuning with a parameter like temperature (T). But unlike its equilibrium counterpart, the *critical state is an attractor of the dynamics in SOC* requiring a separation of time-scales between external driving and internal relaxation (see Bonachela Fajardo 2008). Rather paradoxically, by providing a condition for toppling, the presence of a threshold offers a condition for stability. With a zero threshold, the component sites would be always in an active state, with the system perpetually undergoing avalanches involving many (interacting) sites but little stored energy. At the other extreme (infinite threshold) each site would store the energy received, without interactions or transport of energy; thus making the system undergo unitary sized avalanches. But a non-trivial threshold, plus a conservative rule for redistribution of energy, can lead to correlations between the sites, thus making for a spatially extended response to an external local perturbation. Thanks to closely spaced metastable states, the system can evolve by hopping from one to the other in response to perturbation-triggered avalanches where instantaneous relaxations involving the entire system occur (Bonachela Fajardo 2008).

**Box-I**

Phase transitions; order parameter

• **Phase transitions** were classified by Eherenfest as:
a) First order if there is a discontinuity in the first derivative of the free energy, in the form of a finite energy shift where the order parameter exhibits a discontinuous jump at the transition temperature T with an associated release (or absorption) of latent heat, e.g. as in crystallization
b) Second order if the first derivatives of the free energy—namely the entropy and the magnetization --are continuous (no latent heat) at the critical point, but the second derivatives of the free energy-- namely the specific heat as well as the magnetic susceptibility —show a discontinuity in the form of a divergence (or singularity), as in magnetization of a ferromagnet.
• It was Landau who first introduced a quantitative measure of order appearing at the phase transition, through his definition of an "**order parameter**" (valid at or near equilibrium). It signifies the range over which fluctuations in one region of a system could be affected by those in another. In the case of a ferromagnet, the order parameter is magnetization (M).

This kind of dynamics steadily goads the system towards a state in which the outgoing energy balances the incoming one on average, leading to a scale-free behavior. Unfortunately its meaning remains restricted, by limited consensus (see Halley and Winkler 2008; Turcotte 2001), to the sand-pile model (Bak et al 1987; 1988) whose principal feature is that the (last) `fractal pile'-- symbolizing the critical state -- gets upset by even the addition of an extra grain of sand on top of it due to the local slope of the pile crossing a threshold. This can lead to the toppling of only two grains to an avalanche affecting the entire pile surface with sand-loss at the boundaries, thereby maintaining the stationary critical state (Adami 1995; Bonachela Fajardo 2008; Dickman et al 2000). To generalize to similar phenomena for greater universality, explanations for such “unguided” critical dynamics have been proposed via their implicit association with a tuning parameter (Dickman et al 1998; Sornette et al 1995) like in equilibrium critical phenomena. In an absorbing-state (AS) phase transition, a tuning parameter--the particle density -- determines whether the system is in an active phase (changing in time) or in an inactive phase (stuck in one configuration). The order parameter of these transitions is the density of sites about to topple, called the activity (Dickman et al 1998). *The coupling between order and control parameters helps attract the latter to its critical value* and brings about the phase transition, as well as shows the possibility of a role-reversal (Sornette et al 1995). This makes SOC a plausible candidate among scenarios for long range correlations underlying complex bio-systems (c.f. Anderson’s (1983) spin-glass model). This is since the susceptibility of the organism as a whole (changes in functional patterns manifest in nucleic acid sequence space) to the environment *controlling* its evolution, betrays an intrinsic *memory* mechanism, enabling it to *sense* and *respond* to its *external conditions* by changing its *internal configuration*—via an analogous coupling of control and order parameters. To that end it uses a *diversity of closely-spaced* (metastable) *states*, resulting from *co-operative interactions* between *many d.o.f.s*—all typical ingredients of SOC.

## 2.2 Implications of SOC in life’s emergence

It is interesting to consider a similar control/order parameter coupling-scenario between an environment and its system to understand evolution by natural selection as well as life’s emergence. Indeed, for insights into the major transitions in evolution (Maynard Smith and Szathmáry 1995), leading for instance to improved functionality in an organism, another study (Suki 2012) proposes that phase transitions in the network structure associated with that function can facilitate the transition to improved functions.

Now, computer simulations have provided numerous insights (Kauffman 1993; Kauffman et al 2004; see Gershenson 2010) into the ramifications of lower-scale network parameters on the global dynamics (robustness, evolvability, adaptability). And for insights into complex bio-processes, wherein higher-level behaviour results from interactions at the lower-level, and which cannot be predicted from the latter’s (unit/sub-process) details, it is worthwhile to study systems comprising *non-linearly interacting entities*, i.e. whose state depends on their mutual interactions. Thus, focusing on the nature (inhibiting/activating) of interactions between lower-level units, as well as the network-topology, makes functional bio-networks appear as computing/task-performing devices. Also, network features like modularity, redundancy, and scale-free topology can help the system exploit noise--an asset for functioning in a robust manner despite fluctuations (Fernández and Solé 2004). Furthermore, natural selection may well have exploited such methods to guide the self-organization of genetic regulatory networks towards the critical regime (Gershenson 2010). But this also brings up the intriguing possibility that such networks had themselves emerged via similar tinkering of precedent ones—in a continuous gradual process. More explicitly, we ask if the computing power of organisms that is inherent in the adaptive process (Hartwell et al 1999) could be extrapolated backwards to a rudimentary information processing system in the pre-biotic era that may have guided the evolution of random chemical networks. Indeed Cairns-Smith's (2008) abstraction of control-organization from these computing systems frees them from the material details and helps to extrapolate the LUCA back in time. Here, starting from the pre-biotic era, transitions (c.f. Suki 2012) between information-processing machinery by changing materials/architecture/mechanisms,--in response to environment fluctuations -- require functions associated with the ancestor to be fulfilled by its replacements.

## **3 Liquid crystals (LCs); scaffold paradigm; and bottom-up approaches**.

### 3.1 LC medium as a scaffold-organization

Complex bio-molecules—important components of the control-network—are capable of *large response-effects a la* de Gennes (2005), typical of soft-matter, thanks to correlated motions of their constituent atoms. They display liquid crystalline phases both *in vivo* and *in vitro*. The relevance of an LC medium to biology (see Table–I, adapted from Bisoyi and Kumar 2010), owes it to a feature of cooperativity that facilitates responses to external stimuli (apart from control and stability), but one which is missing in a random mixture of its constituent building blocks (amino-acids, nucleotides, etc.). Besides its intrinsic properties, it can act as an *influential host medium* for the evolution of its embedded materials by controlling their orientation, helping assembly, and transferring its own sensitivity to external-fields due to *steric-coupling* (point 6, Table 1; Sect.5.1), and thus makes it easier to understand Cairns-Smith's (2008) scaffold paradigm. As non-equilibrium states are stable when they act as energy carriers, in the absence of any new functional structures appearing, this medium of cooperatively acting components can offer its own (rudimentary) capacity to act as an energy conduit. Conversely, it can be dispensed with in favour of new emerging structures with superior functions. Thus such dynamic stability ensuing from cooperativity in a medium would have provided time for the *interactions between its randomly engendered materials* to lead to the *gradual* appearance of constructs of *increasingly higher specificity* and *lower connectivity* (c.f. Kauffman 1969), that could range from structures to complex spatio-temporal patterns, capable of canalizing energy more efficiently. This gels with Langton's (1990) emphasis on *the vital dependence of complex computations* requiring diverging correlations in time (for memory), and length (for communications), *on phase transitions*, in the context of life's emergence, by insisting on the primitive functions required for computation, viz., the transmission, storage, and modification of information, so that it can spontaneously emerge as an important factor in the dynamics of a system.

Table –I: The importance of being liquid crystalline.

| | | |
|---|---|---|
| 1. | Capacity to combine order and mobility underlies its crucial role in self-organization and structure formation in biology. | Hamley 2010 |
| 2. | Important biopolymers e.g. lipids, proteins, carbohydrates and nucleic acids display liquid crystalline phases both *in vivo* and *in vitro*. | Hamley 2010 |
| 3. | Like cells LCs can amplify and transmit information | Goodby, et al 2008 |
| 4. | Like cells, they can dynamically respond to a large number of external stimuli e.g. changing chemical concentration, temperature, light, electric, magnetic fields and other environmental changes | Demus et al 1998 |
| 5. | Liquid crystals have potential for electron, ion, molecular transport, besides sensory, catalytic, optical properties | Kato et al 2006 |
| 6. | Control effects: A scaffold medium --as a 'precursor' template *a la* host-level-I --can exert its influence upon its dispersed materials— *a la* guest-level-II (see text). | Bisoyi and Kumar 2011 |
| 6a | Far from inducing distortions various nano-materials dispersed in LC media have been observed to enhance their physical properties. | Hegmann et al 2007 |
| 6b | The anisotropic nature and tenability of LC media can facilitate the alignment and self-assembly of nano-materials *randomly* dispersed within. | Kumar 2007 ; Hegmann et al 2007 |
| 6c | Thanks to the sensitivity of LC media to small external stimuli, the latter can thereby influence the dispersed materials that are *sterically coupled to the host's dynamics*. | Bisoyi and Kumar 2011 |

3.2 The scaffold as a controlled cooperative organization

Rather than suggesting the spontaneous emergence of context-laden biological language from random processes alone, the scaffold-paradigm offers a *pre-existing* environment-responsive functional inorganic *control-organization*-- level-I-- to host/guide the (irreversible) evolution of random organic reactions/assemblies-- level-II. Conceptually, assistance from collective crystal-vibrations (Cairns-Smith 2008) would have elevated the status of a thermodynamically-motivated proto-metabolic process to that of a *function*, while gradual organic 'takeover' of level-I would lead to today's control-network (level-I) *feedback-coupled* with the metabolic-network (level-II), supplying energy and building blocks. Note that in contrast to living systems-- whose ordering source comes from their dissipation of energy (closure; Shapiro 2007), a scaffold awaiting 'takeover' is not constrained to follow this pattern. But it does need a sustained source for its ordering and access to non-equilibrium sources. Now autocatalytic cycles, e.g. reverse citric-acid cycle (Morowitz et al 2000), may have served as disequilibrium-releasing channels besides providing building blocks for the control-network (Copley et al 2005), although they require mechanisms providing *kinetic assistance* and *pruning of side reactions*. Today, regulated enzymes lower activation energy barriers by controlling the orientations of the reactants. True, it is hard to imagine a corresponding variety of enzyme-like specifically-binding surfaces via a crystalline matrix (see Orgel's (2000) perplexity at Wachtershauser's conclusion). Nevertheless, the *effect eventually caused* by the different enzymes, viz. of *trimming the phase-space of the reacting species* (level-II), could have been achieved via the association of some pre-existing control organization --level-I-- with the random pre-biotic reactions.

### 3.3 How cooperativity could complement bottom-up approaches

Approaches considering the emergence of a non-genomic replicator by random drift through autocatalytic closure of simple catalytic molecules before template-replicators (Bollobas and Rasmussen 1989; Dyson 1982; Kauffman 1986; see Hordijk et al 2010) may have overlooked such a 'top-down' pre-existing kinetic principle helping its onset. Besides, a cooperative colloidal system assisting a spontaneous process (function) seems the appropriate medium for supporting/awaiting cooperative phase-transitions in random networks, and selecting gradually emerging ones "taking over" its functions. Now, in looking for the "ultimate ancestors of modern enzymes", Dyson indeed considers the possible role of clay crystals or iron sulphide membranes, but merely as *passively confining* surfaces, which obscures their possible impact on the probabilities of a gradual transition from a random collection of catalytic units to a *co-operative* population, say via the mean-field approximation (c.f., Curie-Weiss model of a ferromagnet), since the population of molecules slowly diffuses over the transition barrier. Nonetheless, taking inspiration from Dyson's (1999) 'cells-first' model, we explore the possibility of a *directed* way to more structured quasi-stationary states --"*possibly with active biochemical cycles and higher rates of metabolism*"-- from within a random and disorganized population of molecules, "*in an assemblage of many droplets existing for a long time*". As mentioned (Sect.1), binding to field-controlled MNPs would have caused a drastic reduction in the phase space available to the reacting organics towards bringing about such a transition thanks to the invisibility of H-fields to organics. It is logical to suppose that magnetic-interactions would restrict the possible orientations of the organic-bound mineral-particle; this physically rules out some interactions/reactions, while kinetically assisting the feasible ones thanks to the flexibility of the magnetic 'template-surfaces' (Ommering 2010; Baudry et al 2006; Sect.5).

As a scaffold hosting random reactions, the field-organized system of nano-particles has potential to fulfill the requirements of distributed-control and kinetic assistance in top-down and bottom-up approaches, respectively, to the origins of life (c.f. Sun 2007). And as the interplay of order and disorder at all scales is also feasible via magnetic d.o.f.s, the emergence of dissipative living systems (c.f. Nicolis and Prigogine 1977) is postulated to have started from such a scaffold-organization dissipating (coherent) field energy for its formation. Although close-to equilibrium initially, over time it got slowly pushed further and further away from equilibrium upon gradual "takeover" by (its selected) organic-based complex components, with an analogous capacity of dissipating homogeneous sources of energy for sustaining their stable and "mutually interdependent dynamics" (Cairns-Smith 2008). This is plausible since the entropy of the super-system---the controlled system plus its environment—would then increase at a faster rate. This field-controlled system offers a mechanism for i) confining adsorbed organics, ii) giving access to diffusing-in ‘food’/materials, iii) permitting generated ‘wastes’ to diffuse out, hence acting like a flow reactor with analogy to Dyson’s pre-biotic “cell”.

## 4 Field-controlled scaffold organization

From among a variety of magnetic effects having implications for life’s emergence the chief emphasis will be on reversible field-induced aggregates to simulate an evolving biosystem. That such aggregates can form (Taketomi 2011) encourages the assumption of their presence in pre-biotic locales, although here one expects greater system-complexity than in the following studies, since there could have been no control on parameters (particle sizes, composition, etc.). But a chief concern is the absence of steric-effects in surface-modified synthetic ferrofluids, to avoid short-range attractive forces. This leaves unaltered action-at-a-distance effects like co-localization of particle-anchored organics, but could affect the scenario of a field-controlled scaffold. Nevertheless, the mutual interplay of magnetic-attraction and charge-repulsion —as in framboid formation (Sect.6)—shows a way to register short-range repulsion between particles.

### 4.1 Brief background

Thanks to thermal fluctuations, magnetic single-domain nano-particles --key players in this scenario--are disoriented at room temperature. A moderate H-field suffices to break the rotational symmetry of such nano- particles, by imposing a directional order against their thermal fluctuations, see Figure 2, taken from Chantrell (1982; see also Klokkenburg et al 2006; Richardi et al 2008). Li et al (2007) describe field-induced aggregates as a phase separation of a particle-concentrated phase from a dilute one. These (close-to-equilibrium) ordered structures --requiring about tens of milli-Tesla fields for their formation-- are dissipative in nature, breaking up when the field is switched off. They are also amenable to control parameters like field strength, sweep rate, concentration, strip- width and strip-thickness. Thus, with the external H- field exceeding a critical value, the original magnetic nanoparticles started to agglomerate into magnetic columns and, with its further increase, formed several levels of ordered structures (Yang et al 2003). As checked by small angle neutron scattering, chain size also depends on the strength of inter-particle interactions (Barrett et al 2011).

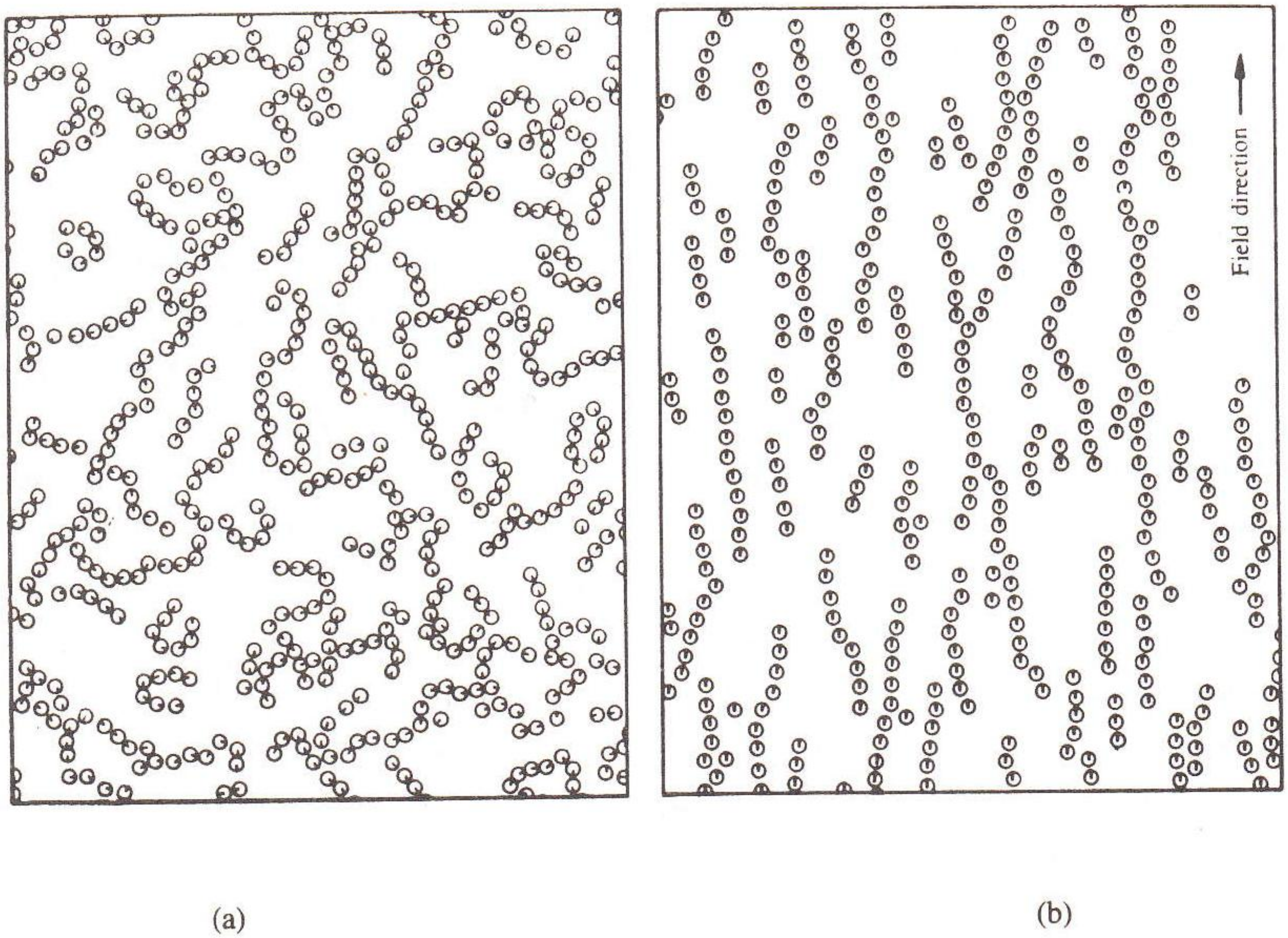

**Figure 2**

An important property of magnetic nano-particles is that of anisotropy (see the classical Stoner-Wohlfarth (1948) model); so that the applied field helps the hysteretic rotation of the magnetization to jump over the magnetic-anisotropy barrier. Next, in general, the relaxation of a single-domain nano-particle can take place via two distinct mechanisms: 1) Brownian- the individual magnetic moments, are rigidly fixed against the nano-particle's crystal lattice so that the particle rotates as a whole; 2) Neel- the individual magnetic moments rotate within the (fixed) nanoparticle. But this would also depend on its physical state. Thus, taking particles whose magnetization is not completely frozen (Neel relaxation time much faster than their measurement time), and dispersing them in a liquid medium would give the colloidal particle's magnetization both Neel and Brownian modes of relaxation. The latter-- proportional to the crystal volume-- characterizes the viscous rotation of the entire particle (irrelevant for dry powders), unlike the former (an exponential function of the volume). Therefore the Brownian mode for return to equilibrium becomes the dominant process for large single-domain particles suspended in a liquid medium. Its characteristic time scale can be studied via ac susceptibility; thus an increase in hydrodynamic radius, such as upon binding to organic ligand --e.g. biotin to avidin-coated nanoparticle (Chung et al 2004)— resulted in a shift in the magnetic susceptibility peak vs frequency curve at 210Hz to 120 Hz. Importantly, the degree of *inter-particle-interactions* (c.f. Mørup et al 2010) can significantly affect this relaxation mode (relevant for further diffusing-in particles into an aggregate). Recall also that over and above the orienting effect of a field, a further enhancement of the magneto-viscous effect (velocity-gradient caused rotation of suspended particles, hindered by field applied perpendicular to sheer flow) was attributed to structure formation (Pop and Odenbach 2006). Furthermore, since the dipolar interaction between two neighbouring particles increases with decrease in intercrystal distance, the particle's aggregation-state should have an effect on the Neel relaxation, due to the dipolar inter-crystal coupling aspect of the anisotropy (Laurent et al 2008).

### 4.2 Analog for confinement

Field-induced (dipolar) interactions offer a ready mechanism for confinement of MNPs by overcoming thermal fluctuations, see Figure 2 (reproduced from Chantrell 1982). The dynamical aggregates, whose components interact via weak reversible complementary dipolar forces, are analogous to living systems with *distributed control* and whose components *dissipate homogeneous sources* of energy. Other than such magnetic field induced aggregation—likely a second-order phase-transition (Taketomi 2011)-- magnetic dispersions can also be ordered by other coherent sources, e.g. light (Köhler and Hoffmann 2003), and electric field (Riley et al 2002; Duan et al 2001). This expands the scope of field-control for access to scaffolds that could have been present in a variety of pre-biotic environments. Now, agglomeration of H-field aligned nano-particles-- dispersed in a fluid-- leads naturally to a bottom-up assembly compliant to top-down control (see Chantrell 1982; Rosensweig 1985), wherein the spread of the aggregate is defined by the field's zone of influence (~ inverse square law). An equilibrium state is reached when the number of particles leaving the aggregate balances those getting attached (Fang et al 2008). Next, we suggest that in an open system, the possibility of further particles diffusing into it and aligning to the assembly "layers" would provide an analog for "replication"/growth.

### 4.3 Correspondence to machine-like components

That bio-systems choose to function near the cooperative transitions of their *myriad different* bio-molecules also gels with 'takeover' from pre-existing modules functioning primitively via collective effects. Bio-molecular machines are many-atom containing molecules whose dynamics seems to be governed by the fluctuation-dissipation theorem (FDT) (Bustamante et al 2005). Their cooperative atomic motions enable reversible switching between conformational states for work cycles. This seems analogous to the capacity of exchange-coupled magnetic moments in an MNP lattice to change their spin orientation in response to local variations of the external H-field (via Zeeman effect).

#### 4.3.1 Diffusion aided processes

Imagine further incoming MNPs, diffusing into *their* field-induced aggregate of MNPs in an aqueous medium (see Figure 2b; c.f. work cycles of a molecular motor moving on a template). Now as a dipole (depicted in blue in Fig.3) diffusively migrates through the 'layers' of the aggregate (depicted in black), in addition to the H-field and bath fluctuations, its orientational state is influenced by the local H-field of its "template" partners forming the aggregate. We also imagine a gentle H-field gradient --stemming from (inhomogeneous) magnetic rocks (Mitra-Delmotte and Mitra 2010a)--that provides both detailed-balance-breaking non-equilibrium as well as asymmetry, to a diffusing magnetic dipole undergoing infinitesimal spin-alignment changes. The gentle gradient-driven diffusion of the migrating dipole (c.f. thermophoresis Duhr and Braun 2006) would thus be periodically perturbed by local H-fields of its ‘template’-partners, leading to alternating low and high-‘template’-affinity states due to the dipole’s magnetic d.o.f., rather analogous to the isothermal release/binding cycles in the priming/operative phases of the molecular machine (Schneider 1991). Within a common FDT framework for asymmetric movements, these changes would be similarly facilitated by thermal excitations from bath, with rectification by either the gentle H-field gradient or the fields of its local ‘template’-partners (see Figure 3 legend). Note also that binding to non-magnetic ligands (e.g. organics) would increase the net potential energy barrier of the particles for interacting with their ‘template’-partners, compared to their ligand-free counterparts. Hence, greater diffusive exploration of the organic-bound particles leads to a bio-molecular motor-like scenario, while the entrapment of the isotropically unshielded ones into an expanding network of dipolar interactions has the appearance of growth phenomena.

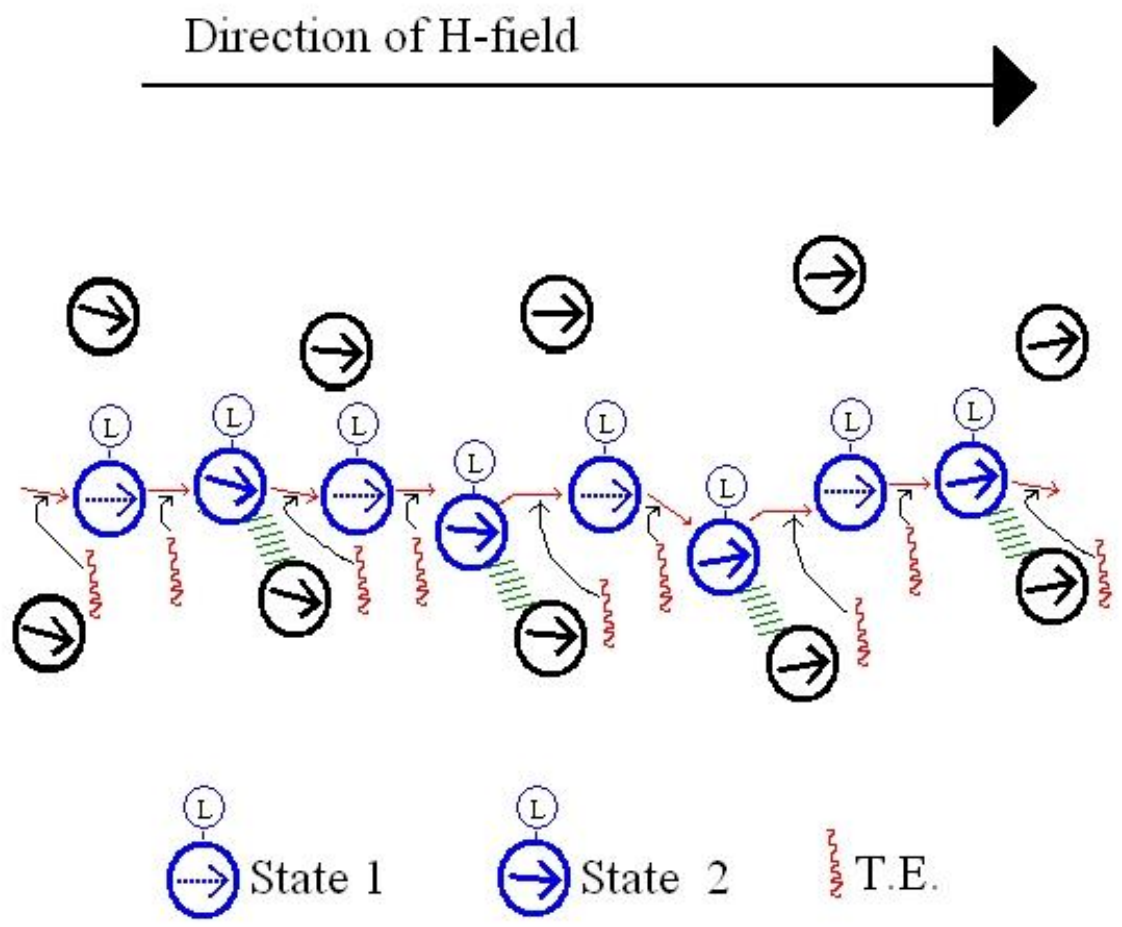

Figure 3

Now, a magnetic ratchet seems promising for the controlled directed transport of micrometer-sized colloids at the solid-liquid interface, as displayed by bio-nano-machines using the ingredients of non-equilibrium source, asymmetry, and a periodically varying potential in space/time. Tierno et al (2008) achieved this on the surface of a ferrite garnet film with a magnetic domain pattern forming a periodic array of stripes with magnetization alternating up and down, and applying time-dependent external magnetic field pulses. Their video-microscopy tracked experiments show the transversal motion of particles on the hard film providing the local ‘template’ fields (Tierno et al 2010). This seems to have potential for being scaled down to nanometer-sized heterogeneities towards a magnetic shift register. Further, tunable heterogeneous field-variations on the nano-scale have not only been used for the controlled movement of aqueous phase dispersed MNPs, but also for their separation based on size of the particles (Tierno et al 2008). The fact that the latter could be used to separate complementary oligonucleotides via a “hot zone” for melting the DNA strands, shows their compatibility with the energy-scales required for controlled biomolecular interactions, and suggests their relevance for envisaged scaffold effects. Also, an interplay of magnetic with micro-convection (Mast and Braun 2010) effects could have potential to cause periodic binding and de-binding between interacting particles.

Plausible mechanisms for ‘organic-takeover’ include the autonomous motion of Janus particles whose surfaces are designed to have asymmetric chemical properties (see Baraban et al 2012). For example, the catalytic action at one end of the particle generates an anisotropic chemical gradient across its surface, and this self-generated force drives the particle’s movement through a liquid medium. Initially during the transition, an external field’s orienting effect may well have guided such directed migration (as in Kline 2005; Gregori et al 2010) before other control mechanisms such as today’s chirality-based ones got installed.

It is important that the size scales of the non-magnetic colloids be kept in mind, when assembling bio-molecules using magnetic effects. For instance, in a magnetizable fluid, large non-magnetic colloids ~ 100nm have been shown to be pulled towards the *lower end* of the field-gradient (exactly opposite to their magnetic counterparts) called negative magnetophoresis (Halverson 2008; Yellen et al 2005)—a method used for their manipulation and assembly by magnetic fields. This volume effect is likely to be negligible for organic ligands, like small peptides considered here; for comparison, a large 20kDa peptide (~170 amino-acids) has an $R_{min}$ of 1.78nm (Erickson 2009).

#### 4.3.2 Magneto-structural transitions

Now, secondary effects of magnetism in a substance are caused by couplings between its different physical properties: magneto-caloric, magneto-electric, magneto-optic, magneto-striction (De Lacheisserie et al 2005), analogously to similarly coupled d.o.f.s (thermal, elastic, electric, etc) of complex biomolecules (Cope 1975). This raises the possibility that similar transitions in magnetic mineral particles (Hemberger et al 2006) comprising field-structured aggregates had assisted some work-cycles, especially since surface-to-volume effects become sizeable at the nanoscale. For example, in the priming step in molecular machine functioning (Schneider 1991), energy is supplied by a field-like (homogeneous) source, typically ATP, plus thermal motions captured from the bath. This is followed by the operating phase wherein dissipative ordering for information gain –recognizing a surface and reducing its conformational uncertainty—and release of entropy to the bath, takes place. The energy-shift via entropy reduction is effectively a *first-order phase-transition*. In the corresponding magnetic scenario for directed transport, an accompanying *magneto-caloric effect* can permit an interchange between system-entropy and bath temperature under isothermal conditions; also a magnetic field-controlled nano-particle assembly mimics recognition-based binding interactions between particle surfaces. Again similar to spatial field inhomogeneities causing motor-like effects, temporal field-variations can cause binding/release cycles between interacting MNPs, analogously to complementary bio-surfaces.

Note that heat released from a reaction, can alter the magnetization of the particles, vide Neel's (1949) study. Further, two analogies of magnetic mechanisms to bio-molecular ones are intriguing: 1) the activation energy of a substrate in a chemical reaction is similar to the anisotropic energy hump of a single domain magnetic nano-particle, flipping from one easy direction to the other; and 2) the interconnections between magnetic elastic and thermal properties in magnetic shape memory materials are rather reminiscent of enzyme dynamics. For example, a change in the material's magnetization by changing an external H-field can not only bring about its deformation (magnetoelastic effect) but also an entropy variation (magnetocaloric); likewise a deformation due to an applied stress, can cause both a magnetization and an entropy change (Giudici 2009; c.f. martensitic-like transformations in cylindrical protein crystals, Olson and Hartman 1982). Alternatively, similar shape-memory effects could also have been effectuated by the diffusive entry of small thermo-responsive polymers, and subsequent binding to magnetically heatable colloids in the scaffolds (Mohr et al 2006; Schmidt 2007; Zheng et al 2009).

### 4.4 Global evolution of aggregates

The field-induced assembly of dispersed nano-particles falls under the general category of granular systems with complex interactions (Aranson and Tsimring 2006), with weak magnetic dipolar interactions providing a global correlation mechanism. The analogy between electric dipolar interaction-based organization in living systems and magnetic dipole interactions in a reversible aggregate (Taketomi 2011) wherein the latter can be influenced by an externally applied H-field, makes them interesting as scaffold-systems a la Cairns-Smith. Ideally, a completely reversible system can capture the interplay between several competing factors, such as magnetic dipolar interactions, thermal fluctuations, screening effects of the medium (Pastor-Satorras and Rubi 2000). Intriguingly, the complex effects of the long-range magnetic dipolar interaction (Huke and Lücke 2004 plus references) —itself dependent on the macroscopic distribution of the particles-- leads to feedback between the external and internally generated fields. This scenario seems to be analogous to the sensitivity of the internal state of living systems to external influences. Although we are unaware of experiments that correspond exactly to these speculations, nevertheless, some insights can be had from the seminal associative memory model of Hopfield extrapolating from physical systems to spontaneous bio-computation as a collective property of autonomously functioning units (Hopfield 1982). Also, the simulations (Ban and Korenivski 2006, Palm and Korenivski 2009) employ a ferrofluid -based associative neural network for pattern storage, wherein inhomogeneous H-fields influence dipole-dipole interactions in the network, with the respective transition probabilities satisfying detailed balance.

In this context, Brevik (2001) first used a life-like system of magnetic floating objects plus thermocycler, as instantiation of uncertainty reduction in producing complementary sequences, and for relating thermodynamics to information—defined as the shared entropy (via patterns) between two independent structures—in living systems. Even without catalysis, spontaneous interactions between monomers bound to a polymer resulted in complementary-string formation in response to environmental temperature fluctuations, thereby demonstrating the self-organization of template-replicating constructs towards Darwinian evolution. Although he used macroscopic objects, this scenario is down-sizeable.

4.5 Far-from-equilibrium regime

Organic bonds (at level-II) could prevent dissociation of field-induced aggregates and enable their drift to locations providing non-equilibrium conditions (c.f. Goubalt et al 2003). For unlike static-field induced equilibrium-like clusters, alternating fields can provide interesting configurations, e.g. dynamical self-healing membranes (Osterman et al 2009), and swimmers (Dreyfus et al 2005). Further, spinning ferromagnetic disks at the liquid-air interface assembled into patterns due to interplay of repulsive hydrodynamic (vortex-vortex) and attractive magnetic (coupling to average field of rotating external bar-magnet) interactions (Grzybowski et al 2000; 2009; Whitesides and Grzybowski 2002). Again, dynamic elongated self-assembled structures-- suspended at the liquid-air interface-- emerged in a certain range of excitation parameters owing to competition between magnetic and hydrodynamic forces. Furthermore, self-propelled “swimmers” formed upon spontaneous symmetry breaking of the self-induced hydrodynamic flows (Snezkho 2011).

Now, the formation of dissipative organic assemblies at level-II requires an energy source, which a scaffold with a *capacity to store* (coherent) energy can support. Indeed, field-tunable aggregates can store polarized (retrievable) light, its wavelength being determined via the refractive index of microcavities formed by the aligned spheres (Patel and Mehta 2011).

4.6 Transfer of heat; electron transmission

Tunable dipole-dipole interactions between MNPs -- via external field strength and its orientation, etc. -- can influence heat *percolation* through the network. Recent results (Philip et al 2008; Shima et al 2009) show a 3-fold enhancement of thermal conductivity of a ferrofluid over the base fluid’s, thus suggesting an efficient percolation mechanism via field-induced aggregation of 3-10nm magnetic particles. Very large conductivity is observed with parallel fields versus low values for the perpendicular mode. Similarly, a field-induced magnetic dipolar network, can also transport (spin-polarized tunneling) electrons (Pu et al 2007). The possibility of percolation of heat and spin-transmission of electrons-- via dipolar interactions-- in a field-induced MNP-network, hosting reactions makes it interesting to consider feedback effects. A reaction at level-II could impact the MNP-network configuration at level-I, say by releasing heat and increasing local temperature or altering the redox state and thereby the magnetic moment of the hosting particle/s (at level-I) (see supplementary information).

Now, thermionic emission via the Richardson effect could have provided single electrons (c.f. pairs from redox reactions) to inorganic-scaffolds, which is interesting in view of the possible role of electron-bifurcation via crossed-over redox potentials in the emergence of metabolism (Nitschke and Russell 2011). Indeed, the gradient-rich mound-scenario studies geological constraints for insights into the emergence of the universally conserved proton-pump--an energy-producing vectorial process (Lane et al 2010; Nitchske and Russell 2009), and where far-from-equilibrium conditions can produce dynamic-cum-catalytic mineral structures (Mielke et al 2011; Sect.6.3). The higher temperature inside the mound could have caused electrons (thermionic emission from alloys) to flow in the direction of the redox gradient. It is interesting to consider the electron passage through field-induced aggregates --expected to be substantial at the gradient boundary-- wherein a reversibly bound particle would suffer a torque effect; this homopolar-motor-like movement may have implications as precursors of rotary motors.

Table-II: Field-controlled colloids for a "scaffold-organization" *a la* Cairns-Smith

| | Field-control assisted 'function' | Living system like characteristics | Speculation based on theory/ reference/s |
|---|---|---|---|
| 1a | Field-controlled aggregates (*c.f. mineral layer sequences in crystal-organization* (Cairns-Smith 1982)). MNP- network configuration susceptible to external influences: size of incoming MNPs, fluxes, H-field, hosted reactions (could change local temperature or MNP's redox state, thus its magnetic moment, etc.); these could impact transport (§4.6). | Confined, environment-susceptible organization; distributed control on independent interacting units; global dynamics irreducible to lower-level components, yet constrained by feedback; closely-spaced configurat-ions (c.f. Anderson 1983); heterogeneity for reaction-diffusion patterns. | Botet et al. 2001; Chantrell et al 1982; Klokkenburg et al 2006; Li et al 2007; Pastor-Satorras and Rubi 2000; Richardi et al 2008; Rosenweig 1985; see Klapp 2005, Sect.4, and references. |
| 1b | Coherent fields (H-field, light, electric field) for alignment, confinement of MNPs into cooperative network; resemble 2nd-order phase- transitions. | Dissipating homogeneous energy sources (ATP) to order components into cooperative organization. | Taketomi 2011; Köhler and Hoffmann 2003; Riley et al 2002; Duan and Luo 2001 |
| 2 | Close-to-equilibrium:<br>i) Weak, reversible dipolar interactions ~ $k_BT$, *sustain* organization in space & time | i) Like weak complementary binding sustains organization in space & replicator in time | Component-level: exchange-coupling in particle-lattice.<br><br>System-level: dipolar-coupling force |
| | ii) external fluctuations can be harnessed at component as well as at system level | ii) fluctuations harnessed by components (work-cycles), and evolving system | |
| 2a | Diffusing-in MNPs aligning & expanding MNP-network | 'template'-aided growth (see (§4.3.1) | Speculation for open system. |
| | Directed transport e.g. nucleotide oligomer-bound MNPs on garnet film) | Ratchet-dynamics of molecular motors (see §4.3.1) | Tierno et al 2008; Tierno et al 2010 |
| 2b | Magneto-structural transitions (like 1st –order) in particle components | *Component-level:* as in work cycles of enzymes, motors. | De Lacheisserie et al 2005, see magnetic materials (§4.3.2) |
| 2c | Associative network (*c.f. varying crystal sequences* Cairns-Smith 1982) in response to external changes | *System-level:* susceptibility to ‘environment’/ *evolution*/analog memory | Hopfield 1982; Huke and Lücke 2004; Palm and Korenivski 2009; |
| 3 | Potential for kinetic assistance in reactions plus trimmed phase-space of bound reactants limits possible reactions (*c.f. “side activity” in crystal paradigm* Cairns-Smith 2008) | Flexible ‘templates’ help juxtaposition of reactants | c.f. Baudry et al 2006 |
| | | Like flow-reactor trimming phase space of bound reactants, curtailing side reactions. | see §1; c.f. Park and Kim 2010 |
| 4 | Far-from-equilibrium: Dynamical structures via alternating H-fields/non-equilibrium conditions. Potential magnon-mode for energy propagation (*c.f. phonons in crystal lattice* (Cairns-Smith 2008) | New self-organized structures like swimmers, self-healing structures, and others not seen in a static field. | Gryzbowski et al 2000, 2009; Osterman et al 2009; Snezhko 2011; Dreyfus et al 2005 |
| | | Field-tunable dispersions can store optical energy (like homogeneous ATP). | Patel and Mehta (2011) |
| 5 | Transfer of heat through aligned aggregate. | Long range energy transfer | Phillip et al 2008; Shima et al 2009 |
| 6a | Transfer of electrons (spin-polarized) through aggregate | Long range electron tunneling | Pu et al 2007 |
| 6b | *Field-aligned* aggregate for spin-transmission (above) | *Chiral* assemblies for selective spin-transmission | Naaman and Zager 2011 |
| | Magneto-optical properties: field-induced birefringence; Faraday rotation, ellipticity; linear, circular dichroism | Analogous to properties of biological matter | Davies and Llewelyn 1980 |
| 6c | Current carrying particle *a la* homo-polar motor, §4.6 | Vectorial proton-transfer for torque in rotary motor | due to Lorentz force |
| 7 | Effect of H/electric-fields on MNP/thread suspensions | Resemblance to mitotic spindle | see Rosenberg et al 1965 |
| 8 | Merger of magnetic assemblies from different locales | Horizontal information/gene transfer | -- |

## 5 Towards cooperative transitions

In general, depletion forces (Asakura and Oosawa 1958; Marenduzzo et al 2006) can enable aggregation in a given location provided there is an excess above a critical concentration of interacting particles due to specific binding or anisotropic forces. Above a threshold concentration, packing (translational) entropy stemming from shape anisotropy (causing decrease in orientational at the cost of positional entropy) could help overcome rotational entropy and break orientation symmetry, thus maximizing total entropy (Onsager 1949; Dogic and Fraden 2005). This route could have been accessible to rod-like mineral colloids (Davidson and Gabriel 2005; see Hamley 2003). Indeed, mineral liquid crystals (Gabriel and Davidson 2003; Lemaire et al 2002; Vroege et al 2006; van den Pol et al 2010) appear promising as "readily available" candidates with potential to provide a cooperative medium with sensitivity to environmental stimuli (Cairns-Smith 2008), and this calls for a database of such minerals in prebiotic locales. The route to LC phases has also been attempted directly from a mixture of organics using entropic forces for achieving self-assembly. Nakata et al (2007) elegantly demonstrated the assembly of short complementary double stranded DNA into LC aggregates. The unpaired oligomers maximize their entropy via phase-separation of the rigid duplex-forming oligomers into LC droplets (minimizing their volume). As mentioned earlier (Sect.1, 3), to serve as effective conduits for energy flow the components of aggregates en-route to life need to bind via weak cooperative interactions (c.f. weak/reversible and transient yet specific complementary interactions enable execution of bio-functions). Now we shall briefly review some experiments to explore the potential of field-controlled aggregates to influence the phase-space of their organic guests, noting that *alignment, complementary binding interactions, and homo-chirality* are important requirements towards decreasing the excluded volume of packed molecules as in liquid crystalline phases (Table-I).

### 5.1 Alignment/orientations of mineral-anchored organics

Field-aligned particles seem equipped for the scaffold requirement of influencing their guest particles by transferring their externally-induced orienting ability to their anchored organics. We imagine that in locations enriched in interacting organics (see below), transitions in abiogenic polydisperse organics to LC-phases could have been aided via coupling of their orientations with those of 'doping' low volume concentrations of external field-aligned ferromagnetic particles ( *a la* "ferronematic" phases coined by Brochard and de Gennes 1970), which could have increased the effective susceptibility of the fledgling organic LCs. This decrease in the effective magnetic Frederiks threshold could have led to their alignment and fractionation in the presence of weak H-fields. Moreover, recent work (Podoliak et al 2011) suggests that although ferromagnetic particles induce a low-field response, the intrinsic diamagnetic susceptibility of the ferronematic comes to dominate its magnetic response behaviour-- a scenario intriguingly reminiscent of Cairns-Smith's 'organic take-over'.

### 5.2 Increasing co-localization of interacting organic pairs

As abiogenic organics were unlikely to possess shape anisotropy, *a high concentration of complementary binding pairs with specific interactions* would have been crucial for the formation of LC phases. Indeed, for reasonable probabilities of collective transitions from disordered to ordered mutually catalytic ensembles, the ingredients required are simply stable and confined populations of molecules, whereby chance discrimination of *specific interactions* could bring about catalysis; and increasing number of such mutual interactions eventually causing catalytic reproduction of the whole set (Dyson 1999; Kauffman 1993). Now, Hunding et al (2006) propose that a web of aggregates resulting from selection and growth by complementary-binding between diverse pairs of molecules across pre-biotic locales, can explain the emergence of specific interactions between like and unlike molecules in life-processes. But abiogenics could have been present in diluted solutions as well as high local concentrations via physical mechanisms (Budin and Szostak 2010). To increase the concentration of complementary organic pairs in a given location, consider a possible magnetic d.o.f. of the organic-bound mineral particles (Sect.1). In contrast to specific binding interactions, non-specific binding in concentrated media can be overcome by magnetic forces, thus offering a way to select pairs with binding capacity above a threshold (see Ommering 2010). We suggest that, thanks to a field's 'action at a distance' capacity, its responsive particles—in the event of being bound to one of a pair of complementary-binding organics – have the potential to 1) aid the pairs to find each other by facilitating their detection in dilute to concentrated media (e.g. Pan et al 2012); and 2) *chaperone* the recognition process to assist their binding (e.g. Baudry et al 2006) thanks to flexibility of the colloidal field-aligned 'templates'. Baudry et al (2006) demonstrate how one-dimensional confinement of magnetic colloids in the presence of an H-field considerably accelerates the recognition rate between grafted receptors and their ligands, as measured by turbidometric detection of complexes in the absence of the field. They suggest that since confinement significantly augments the colliding frequency, the same also causes a large increase in the attempt frequency of recognition. An extension of such experiments by first feeding the (open) system with a slow input of nano-particles chemically conjugated to moieties like nucleotides/small peptides-- and consequently checking for the incorporation of labeled complementary units-- could be done in the absence/presence of an applied moderate H-field. Figures 4a and 4b reproduce the experiments by Slater's group (Ho et al 2009) who have used magnetic templates to adhere magnetically labeled cells, to illustrate how a local field, say from magnetic rocks (Sect.6) could have influenced the dynamics of magnetic particle-anchored organics.

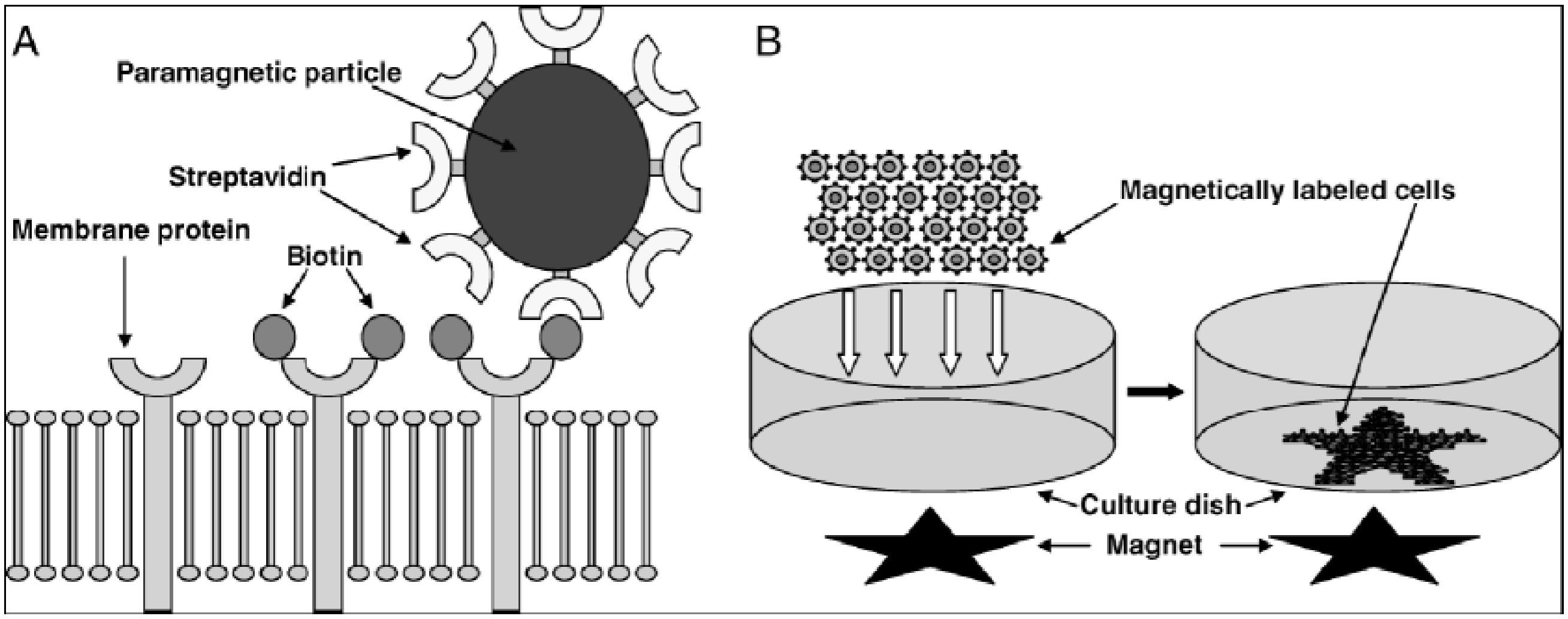

**Figure 4a**

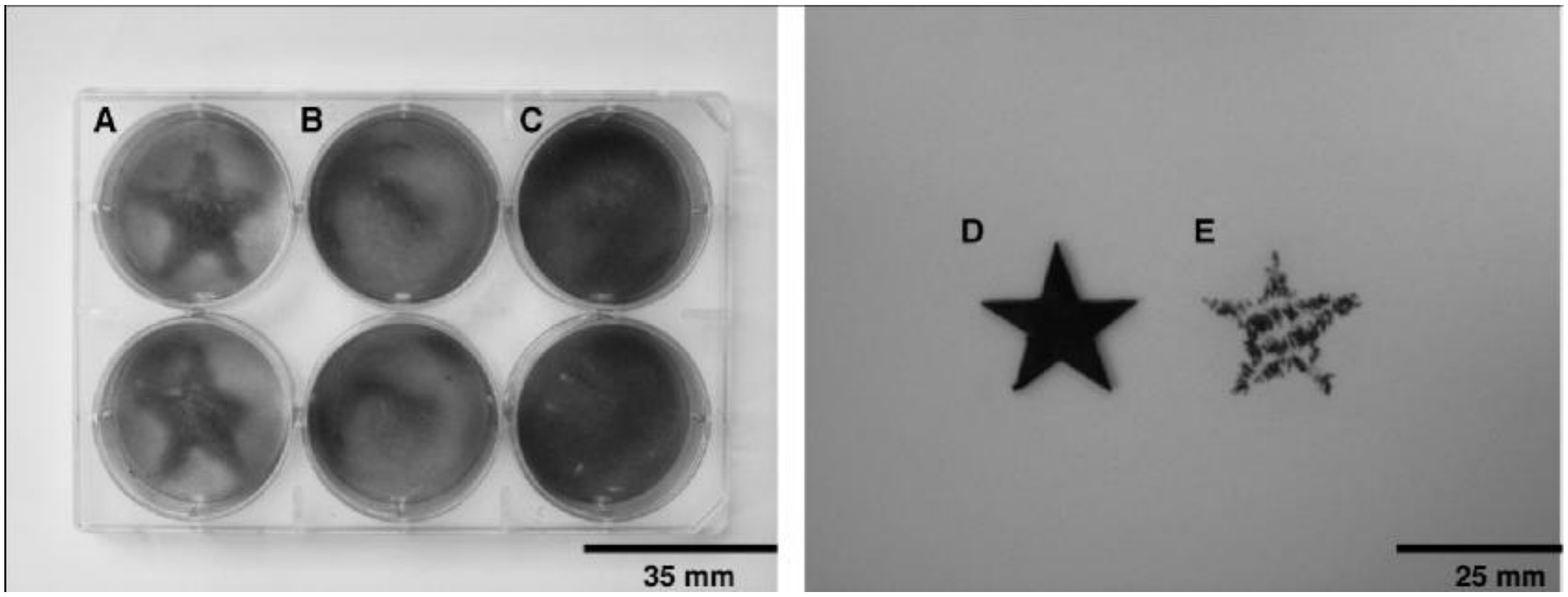

**Figure 4b**

5.3 Homo-chirality

Perhaps the most intriguing implication of a role of magnetic-fields in life's emergence comes from the *homo-chiral* nature of its building blocks that respond differently to left/right circularly-polarized light. Indeed, the findings (Carmeli et al; Naaman and Zager 2011) have further fuelled this speculation by relating the capacity of scaled up chiral assemblies of these building blocks to selectively transmit electrons according to their spin-polarization. To explain in slightly more detail, Rosenfeld's (Wagniere 2007) chiral 'rotational strength' parameter m.d (a unique combination of P, T violating joint PT conservation !) brings out the role of the magnetic part of the e.m. field in 'twisting/reorienting' the magnetic moment about the field axis. Its interaction with the electrical e.m.f. component causing electronic orbital transitions (by polarizing electron cloud across the molecule) thus leads to transitions in a chiral molecule, with d and m components being parallel and antiparallel, respectively, hence averaging out to zero for a symmetric molecule. Feedback makes these interactions complex since oscillating H-fields can cause charges to move and vice versa. A moving charge in turn affects the properties of the carrier transporting it, and electrons have both charge and spin. Just as an electron's response to an electric or magnetic field shows up as a translation (via charge) or rotation (via spin), likewise the response of its carrier to the electric and magnetic fields is one of charge-translation or spin- rotation, respectively. In this context recall the hypothesis of Garay et al (1973), viz. the electron magnetic moment and the magnetic transition moment of the electronically excited chiral molecules could interact. Thus, the magnetic transition dipole could influence the probability of the triplet state of the optically active molecules, electron transport, and stereo-selectivity. Now, the findings of Naaman's group on orienting effects of weak H-fields on bio-membranes, suggest that spin-transmission in the scaled-up versions of the chiral building blocks follow analogous rules of magnetic interaction to those of the individual building blocks. They reported *unexpectedly high* selectivity in transmission of spin-polarized electrons that are consistent with giant magneto spin selectivity in inorganic magnetic films and related colossal magneto-resistance effects. Here charge-transfer from metal substrate converted adsorbed chiral bio-molecular layers from electric to magnetic dipoles, due to *cooperative effects*. Charge redistribution leads to altered electronic structure via unpaired electrons on adsorbed molecules, rendering them paramagnetic. Although spin-filtering effects are achieved in spintronics by applying an external field to induce magnetization in ferromagnetic thin films, magnetization in their bio-counterpart, i.e. a layered assembly of dipolar chiral molecules, is based on two stages: 1) the H-field-created by transfer of charge (electron/hole) through chiral molecules aligns the magnetic dipole of the charge transferred; 2) subsequent exchange interactions in the layered domain keeps them aligned.

These observations by Naaman's group link up two seemingly unrelated aspects of homo-chiral biological units, viz. *selective spin-transmission by their scaled-up assembled versions*. It is gratifying to note that this irreducible picture can be roughly met via H-field aligned colloids (Pu et al 2007; Sect.4.6). Next, consider the latter as hosts of a proto-metabolic reaction: the picture of high energy electrons transferred to sinks like $CO_2$ (Trefil et al 2009) at level-II, seems consistent with that of functional 'takeover' (c.f. Cairns-Smith 2008) of the (spin-polarized) electron carriers at level-I by chiral organic assemblies, whose formation from building blocks would cause phase-space reduction. This is since chiral asymmetric structures, such as helices, provide further scope of entropic interaction-driven phase transitions: the excluded volume decreases in going from packing parallel helices that are out-of-phase to in-phase ones, and at an angle to facilitate interpenetration into each other's chiral grooves (Barry et al 2006). As to the source of engendered building blocks-- again in agreement with the two-level scaffold paradigm -- note that field-controlled aggregates also have the potential to host the formation of chiral organic guests. Although using a different source, Rosenberg (2011) has demonstrated that substantial chiral-specific chemistry was induced by spin-polarized electrons which were provided by radiating the magnetic substrate, adsorbing the chiral organics, by an ionizing source. (The spin dependence of DOS near the Fermi energy in magnetic matter suggests how the colloids could act as spin filters). This is apart from the implications of field-controllable particles in asymmetric chemical synthesis (Ranganath and Glorius 2011), and controlling chemical reactivity via spins (Buchachenko 2000).

## 6 Magnetic framboids; the mound scenario; a fractal scaffold

### 6.1 Framboids and fractal framboids

As a possible scenario towards realizing a field-controlled scaffold, we briefly look at framboids, whose raspberry-patterns inspire their nomenclature. A number of structurally different minerals other than pyrite, i.e. copper and zinc sulphides, greigite, magnetite, magnesioferrite, hematite, goethite, garnet, dolomite, opal, and even in phosphoric derivatives of allophone (Sawlowicz 2000)—form framboids, suggesting *a physical mechanism of formation*. Their formation is a dynamical self-organizing process: The nucleation of a supersaturated solution by the first-formed crystal triggers the separation of many crystals of the same size. Their ordering is an outcome of the interplay of close-packing attractive (such as surface-tension, magnetic) and repulsive (e.g. electric) forces (see Sawlowicz 2000). Next, studying their presence in sedimentary environments, Sawlowicz (1993) found framboids to be structured over a hierarchy of three size-scales: microframboids, to framboids, to polyframboids; he suggested the formation of nano-framboids, comprising microcluster aggregations (~ 100atoms), by analogy with the 3-scale framboidal hierarchy.   Pictures of polyframboids and aggregations of minute particles forming spherical grains (microframboids) in framboid are reproduced in Figure 5, from Sawlowicz (1993). Based on observations, Sawlowicz proposed a formation mechanism by which the original super-saturated gel-droplet would undergo subsequent divisions into immiscible smaller droplets; further subdivisions would depend on a number of factors (e.g. initial size, iron concentration, gel stabilization, viscosity, activity of sulphur species), wherein a key role is played by the colloid-gel phase in leading to the fractal forms. Also, the exclusion of organic compounds led to simple framboid formation via an aggregation mechanism, while in experiments with organic substance stabilized gel-droplets, fractal framboids formed by particulation.

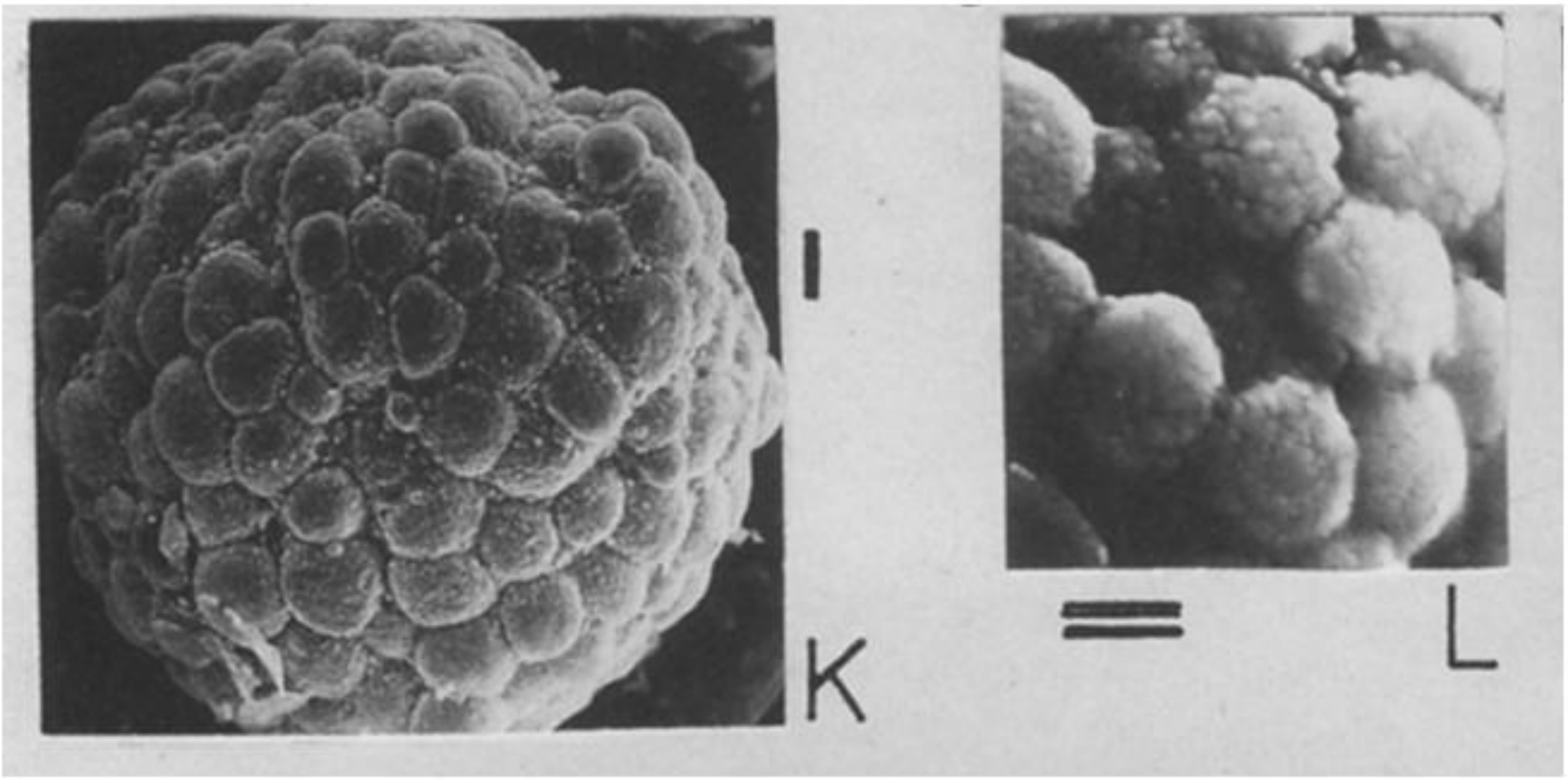

**Figure 5**

6.2.1 Mineral 'relics'

Besides having a striking resemblance to FeS clusters in ancient enzymes (Sect.6.3), the mineral greigite has magnetic properties. Now today's enzymes control electron transfers in FeS clusters (Noodleman et al 1995; 2002) exploiting their sensitivity to local micro-environment fields (organic ligand, solvent, etc). This gels with the picture of controlling enzymes 'taking over' from functioning catalytic-cum-magnetic components of a field-controlled network. Further, from observations of (bio-mineralized) fractal greigite framboids (Preisinger and Aslanian 2004), it seems to be compatible with a nested organization; it can also be found in the magnetosomes of many bacteria (Reitner et al 2005; Simmons et al 2006). Indeed, magnetic mechanisms are hardly "unfamiliar" to living systems, being present across the kingdoms, and evolved at different times (Kirschvink  and Hagadorn 2000; Posfai et al 2001; Winklhofer and Kirschvink 2010).

### 6.2.2 Wilkin and Barnes model

Wilkin and Barnes (1997) have explained the formation/stability of micro-meter sized pyrite framboids, using an interplay of negatively charged repulsive and magnetically attractive forces (in precursor greigite), where a size > 100nm would orient crystals to the weak geo-magnetic field ~ 70 microTesla. Assuming a spherical geometry, the critical grain diameter of constituent crystallites comprising the framboid interior $d_c = 2a$, where $a > 1$, is given by $d_c = (6k_BT/ \mu_0\pi M_{sat}|H|)^{1/3}$ . This result can be obtained from the inequality $W_{WB} > k_BT$ where we define $W_{WB} \equiv \mu_0 M_{sat} V H$. Here $k_B$ is Boltzmann's constant and $\mu_0$ the permeability of vacuum. When aligned parallel to the weak geomagnetic field (~ 70μT), $d_c$ = 0.1 μm. [Ferrimagnetic greigite has a saturation magnetization value $M_{sat}$ at 298K ranging between 110 and 130 kA/m. On the basis of microscopic observations by Hoffmann (1992) of natural greigite crystals, single-domain particles are roughly less than a micrometer in size].

Now for an extension of this field-assembly mechanism to the nano-scale, an extrapolation using the above formula for $d_c$ shows that an H-field for accreting 10nm sized particles--as for ferrofluids-- would have to be ~1000-fold stronger than the weak geo-magnetic field. And as there was no trace of any geo-magnetic field at ~ 4.1-4.2 Ga (Hazen et al 2008), the time when Life is believed to have been already initiated (4.2-4.3 Ga) (Russell and Hall 1997; 2006), we need extra-terrestrial sources, eg. *meteoritic matter*, for providing local H-fields (for e.g. see Sect.6.3 (next), Mitra-Delmotte and Mitra 2010a).

### 6.3 The hydrothermal alkaline mound scenario

A colloid-gel environment in the Hadean with potential for magnetically formed framboids (Mielke et al 2011) is the alkaline seepage site mound scenario (Russell et al 1994; see Sect.4.6), wherein greigite ($Fe_3S_4$) provides the 'continuity' link to iron-sulphur clusters (Sect.6.2.1). Briefly (see Figure 6 reproduced from Russell and Martin (2004), and Russell and Hall (2006)), water percolating down through cracks in the hot ocean crusts would react exothermically with ferrous iron minerals, and return in convective updrafts infused with $H_2$, $NH_3$, $HCOO^-$, $HS^-$, $CH^-_3$ ; this fluid (pH ~ 10 ≤ 120 C) would exhale into $CO_2$, $Fe^{2+}$ bearing ocean waters (pH ~ 5.5, ≤ 20 C), and create porous mounds consisting of brucite, Mg-rich clays, carbonates, Fe-Ni sulphide and green rust-- self-restoring reactors for titrating the hydrothermal fluid with the sea-water (Russell and Arndt 2005)-- towards reducing $CO_2$ (Russell et al 2005). Despite the low levels of bisulphide in alkaline solutions, (Mielke et al. 2010) have shown the potential of the hydrothermal solution to dissolve sulphydryl ions from sulphides in the crust that are expected to flow over ~30,000 years-- fulfilling the continuity of conditions required for abiogenesis. Here, the ensuing super-saturation in response to gradients (stark contrast of pH, temperature, etc.) would spontaneously result in colloidal precipitates of FeS (amongst other compounds, e.g. traces of W, Mo); these barriers would obstruct further mixing of the solutions, leading to the creation of non-equilibrium gradients (pH, redox, temperature; see Sect. 4.5-4.6) across these catalytic membranes, growing by hydrothermal inflation. And, abiogenic molecules (corresponding to metabolic/control levels) would coordinate with each other (Milner-White and Russell 2010; 2011) in inorganic compartments and dynamically ordered framboidal reaction sacs (Russell et al 1989).

Indeed, spherical, ordered aggregates of framboidal pyrite (~ 5μm diameter) were found in fossil hydrothermal chimneys (Boyce et al. 1983; Larter et al. 1981; see Figure 7 provided by Boyce (PhD. Thesis, 1990). Further, Russell et al (1990) have noted the size similarities between magnetosome crystals and pyrite crystallites (~ 100nm in diameter) comprising the interior of framboids that seemed to have grown inorganically from the spherical shells of iron-sulphide gel. And, it is gratifying to see laboratory-formed membranes under non-equilibrium conditions revealing globular clusters that comprise or are attached to, the inner walls consisting of mackinawite and greigite (Mielke et al 2011). These clusters (~1–10 micrometer diameter) resembling framboids, appeared similar to those in the fossilized chimneys, while the outermost crystalline layers were primarily composed of ferrous hydroxide [$Fe(OH)_2$] with an admixture of nanocrystalline mackinawite; the latter were located where the highly alkaline flow could have intercepted the ferrous iron-bearing fluid, and the former where the acidulous iron-bearing solutions could access the alkaline interior of the chimneys walls with concomitant precipitation of the framboids.

### 6.3.1 Extension of mound scenario

Note that negatively-charged mineral greigite forming under mound conditions, where pH is well above 3 (Wilkins and Barnes 1997), resembles an aqueous-based ferrofluid. Significantly, the key to stabilizing its colloidal-gel state lies with organics (Rickard et al 2001). The formation of colloidal magnetic minerals like greigite in the mound scenario makes it relevant to look for a control mechanism via an H-field, such as provided by rocks at the base of the mound. Primary magnetism is plausible via extraterrestrial meteoritic particles (unpublished work of Ostro and Russell; see Mitra-Delmotte and Mitra 2010a). And, this is expected to be reinforced by secondary magnetism thanks to serpentinization and production of magnetite. Magnetic networks can also bring together mechanisms harnessing different gradients via further colloidal/mineral precipitates enveloping the mound (c.f. H-field-influenced growth pattern of precipitated tubular structures, Stone and Goldstein 2004).

We saw (above) that the formation of precipitates leads to progressive growth of the chimneys: their growing front is soft and gel-like, whereas the chimney parts lower down harden as a result of aging. The progressive precipitation of colloidal particles containing magnetic components could have led to detrital remanent magnetism in the chimneys, thanks to the magnetic rock-field at the base of the mound, causing the physical alignment of the magnetic particles at the time of deposition. Thus chimneys/dendrites comprising magnetic minerals, and growing as a result of slower diffusion-aided processes, suggest that further magnetic ramifications such as spin-effects may have occurred within the thermal gels at the soft growing chimney front. Also, fractal

aggregates—dendrites, framboids, etc.—show the possibility of reduction to lower size scales, and of being controlled by external fields (Botet et al 2001; c.f. electric-field, Tan et al 2000).

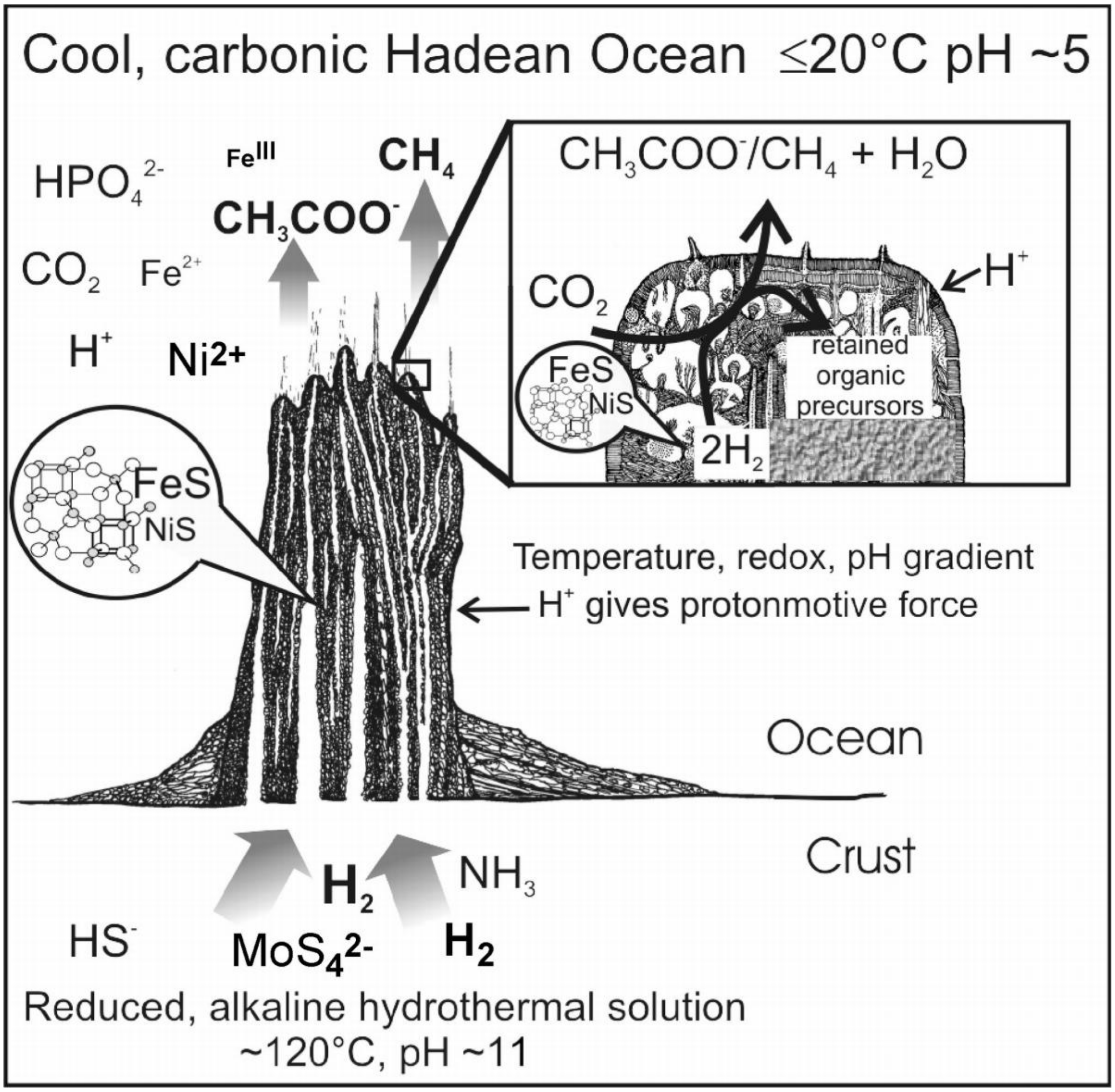

**Figure 6**

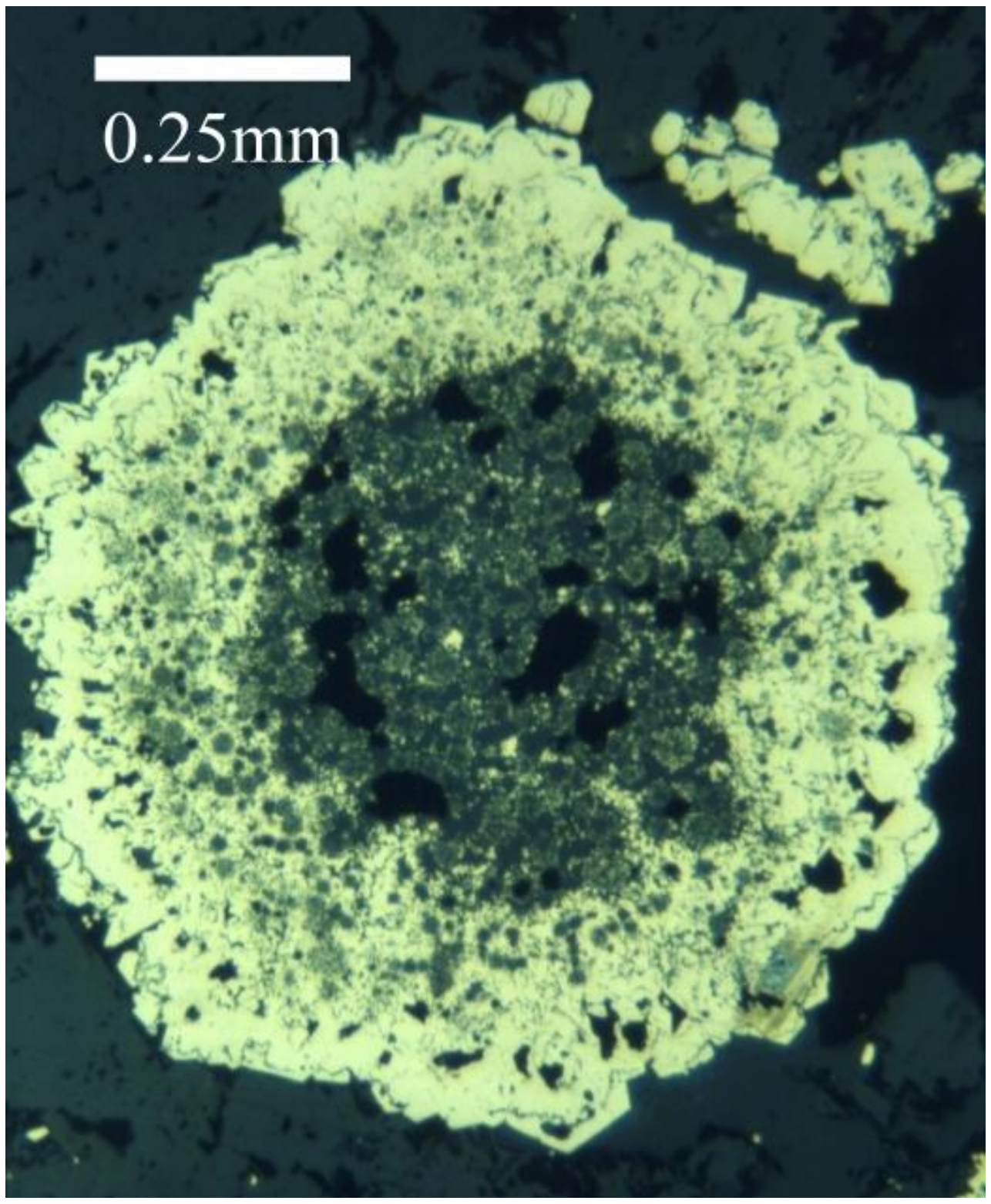

**Figure 7**

6.4 Fractal-network: inorganic scaffold

The influence of network topology on its properties has attracted interest (Albert et al. 2000; Aldana and Cluzel 2003; Amaral et al. 2000)-- e.g. robustness via scale-free networks, fast communication through small world networks, etc.-- although no consensus exists on its relationship to biology (Khanin and Wit 2006). While ubiquitous fractal patterns in biology at the controlling level-I are likely to be the fruits of evolution and selection, in life's origins such *nested architecture* could have been accessed via (host) inorganic scaffolds assisting/controlling guest-level-II processes. Fractals (West and Goldberger 1987) have been noted for their capacity for "*Fitting nearly infinite networks into finite spaces*" (Onaral and Cammarota 2000). Indeed, a nested organization in biosystems permits processes to operate locally at equilibrium despite the whole system/subsystem maintaining itself far-from-equilibrium (Ho 1997). Further, these dynamical patterns are realized via reversible gel-sol transitions, using the capacity of living systems to exist at the boundary of solid and liquid states (Trevors and Pollack 2005). Since field-induced (dipolar) ordering offers an interaction mechanism that does not make use of any chemical or geometrical constraints of the particles, we speculate that this would enable the independently acting components to explore structural configurations at every scale. And, inspired by the observations of Russell et al (1989; 1990), Sawlowicz (1993; 2000), and Preisinger and Aslanian (2004), we have conjectured that moderate local magnetic fields could cause nested formations at the nano-scale as soft scaffolds for life's emergence (see Merali 2007; Mitra and Mitra-Delmotte 2011; Mitra-Delmotte and Mitra 2010a; 2010b; 2012). Further, in gradient-rich (redox, pH, temperature) environments, as in the mound, gradient-dissipating organic fractal structures (Seely and Macklem 2012)-- assembling from building blocks at level-II-- could have gradually replaced the functioning modules of the control-level-I inorganic networks. The tunability of inter-particle distances in the colloidal networks (above) via an H-field (and influencing percolation of heat and electrons (Sect.4.6)), also suggests a route for modulating the *connectivity* of organic networks (Kauffman 1993), the former providing an underlying manifold for guiding (c.f. Gershenson 2010) the organization of the latter. Furthermore, a heterogeneous inorganic organization can assist reaction-diffusion patterns (Turing 1952; Kopelman 1989; Russell et al 1994).

## 7 Conclusions and scope

LC assemblies can be regarded as the minimal units of living systems sharing their environment-response behaviour that can be traced to cooperative interactions. Next, a simplified 2-tier projection of living systems shows the interdependence between the metabolic network (level-II) and the control network of complex biomolecules with LC properties (level-I). Extrapolating this scenario to life's origins, shows that macroscopic energy flow in the metabolic reaction cycles at level-II can be mapped to that in similar attractor cycles in pre-biotic locales. But no corresponding organic equivalents seem to be available for the control network (level-I), with microscopic energy transfers, and which lower kinetic barriers and catalyze level-II reactions. To that end, Cairns-Smith's crystal-scaffold-- a level-I organization-- is extended to field-responsive mineral particles, since the intermediate regime between diffusion-limited and field-driven aggregation of anisotropic colloids seems capable of accessing the features of scaling and controlled mobility in disordered liquid medium. Such a cooperative manifold of reversible interactions achieved via coherent sources enables confinement (solid-phase-like), yet allows random sets of (MNP-bound) organics to interact (liquid-phase-like). Further, this LC-like cooperative organization is susceptible to external influences (size and magnetic moment of incoming MNPs, fluxes, etc) that can change its function-associated configuration, leading to feedback between guest and host levels. A function—of assisting a spontaneous process—associated with an organizational "whole" corresponds to the anatomy of bio-networks, and induces selection of the functional configuration. Again, via this susceptible configuration, the inorganic network can influence the evolution (irreversible) of its sterically-coupled organic guests (level-II) and cause their mutual coupling, say, via an SOC-like mechanism (among those for generating long range correlations). We speculate that the capacity to act as a low resistance channel of energy flow would have been a pre-requisite for a long range correlation scenario, towards becoming a computing system. Moreover its influence on the phase-space of its associated organics (Sect.5) would have oriented their assembly and dynamics towards a kinetic (Pross 2005) direction (breaking free from thermodynamic constraints). This would have poised the system for a series of phase-transitions with appropriate replacements "taking-over" the sustenance and continuity of its functions, till achievement of closure and life's emergence. We hope the testable ideas presented here will motivate further research.

## Acknowledgements

We are grateful to Prof. Michael Russell for inspiration and support (data, figures, key references). We thank him and Kirt Robinson for bringing the work of the Naaman's group to our notice. For kind permission to reproduce their work we thank Dr. Adrian Boyce (labeled framboid pictures); Prof. Roy Chantrell (simulation of field-induced ordering in ferrofluids); Prof.Nigel Slater (magnetic cell patterning); and Prof. Z. Sawlowicz, (framboid pictures). We are grateful to the Reviewers for valuable suggestions and references, (e.g. Seely and Macklem 2012). GM-D is grateful to organizers and participants of the LIO Spring School for discussions, especially to Prof.Eors Szathmary for his criticism and suggestions on presentation. We thank Dr. B.M. Sodermark for suggestions (fractal structures); Prof. Anand K Bachhawat for encouragement; Gaetan Delmotte, Clarisse Grand for Hama-bead patterns (Figure 1); and Mr.Guy Delmotte for computer support. We are grateful to Dr. Jean-Jacques Delmotte for providing financial and infrastructural support.

**Figure legends:**

**Figure 1:** Towards facilitating the evolution of organic reactions/interactions (guest level-II) via a controlled inorganic scaffold (host- level-I) *a la* Cairns-Smith: a) The probability of forming complex stable dynamical patterns decreases with increasing number of organic molecules. This can be aided via selection by a pre-existing functioning organization—the crystal-scaffold or level-I (represented by a white pin board) acting as 'traps' for functioning assembled modules from level-II (represented by a 'bottom-up assembly' of coloured beads). For eg., a variety of recognition-like interactions between organic 'building blocks' are required (not all are shown) to construct the unit leading up to the four-fold symmetric structure. Shown on top is the new organic organization which has functionally replaced the original crystal one at level-I. b) To make this scenario compatible with soft colloidal dynamics and facilitate the 'takeover' of level-I by a hierarchy of functioning modules, we suggest a reversible field-stabilized scaffold with a modular organization—represented by a transparent pin board. A stable inorganic scaffold is also compatible with the simultaneous emergence of (and replacement by) different types of organic spatio-temporal correlations, and as each of these would be dependent on the scaffold, any external tinkering with the latter's d.o.f.s, would also impact the different organic networks and *facilitate their mutual coupling* (see text).

**Figure 2:** Monte Carlos simulation in 2D: (a) clustering without H-field; (b) chaining under H-field (reproduced with kind permission from Chantrell et al 1982; see also Rosensweig 1985).

**Figure 3:** Speculated asymmetric interactive diffusion of further incoming ligand (L)-bound magnetic-nanoparticles (MNPs), represented in blue, through a field-induced MNP aggregate, represented in black (in aqueous medium) in response to a gentle gradient (non-homogeneous rock field). State 1/ State 2: lower/higher template-affinity states of the ligand (L) -bound MNP, in blue; green lines signify alignment in State 2; T.E. or thermal energy from bath; rock H-field direction indicated on top (Mitra-Delmotte and Mitra 2010a). A spatially non-homogeneous H-field is imagined (via magnetic rocks) that provides both detailed-balance breaking non-equilibrium and asymmetry, to a diffusing magnetic dipole undergoing infinitesimal spin-alignment changes. In addition to the external field and the bath fluctuations, its orientational state is influenced by the local H-fields of its "template" partners (forming the aggregate) that would periodically perturb its directed diffusion. This would lead to alternating low and high-'template'-affinity states due to the dipole's magnetic d.o.f., analogous to the isothermal release and binding cycles of the molecular machines on nucleic acid/protein templates, respectively. These changes would be similarly facilitated by thermal excitations from bath, with rectification by either the gentle H-field gradient or local template-partner H-fields (see text).

**Figure 4** Patterning of magnetically labeled cells by Slater and coworkers (Ho et al 2009): (a) Schematics of the procedure for magnetic cell labeling and patterning. A: Magnetic cell labeling. Cell membrane proteins were first biotinylated and subsequently labeled with streptavidin paramagnetic particles. B: Magnetic cell patterning. A star-

shaped magnet was attached under the culture dish. Magnetically labeled cells were added and patterned onto the plate by the magnetic field. (b) Magnetic cell patterning of biotinylated human monocytes (HMs) labeled with streptavidin paramagnetic particles. A: Magnetically labeled HMs were successfully patterned by the star-shaped magnetic template. B: Magnetically labeled HMs were not patterned in the absence of the magnetic template. C: The non-labeled biotinylated HMs were patterned unsuccessfully by the magnetic template. D: Original magnetic template used to pattern HMs. E: Magnetic field profile of the magnetic star template used, as visualized by using iron filings to locate magnetic field maxima. Figures and legends taken from Ho et al (2009) with kind permission from Prof.Nigel Slater; "Copyright (2009) Royal Society of Medicine Press, UK".

**Figure 5:** Development stages of pyrite framboids: scanning electron microscope image of (K) polyframboid; (L) aggregations of minute particles forming spherical grains (microframboids) in framboid; pictures reproduced from Sawlowicz (1993) with kind permission (single bar = 7 micrometer, double bar = 0.5 micrometer).

**Figure 6. The hydrothermal mound as an acetate and methane generator**
Steep physicochemical gradients are focused at the margin of the mound. The inset (cross section of the surface) illustrates the sites where anionic organic molecules are produced, constrained, react, and automatically organize to emerge as protolife (from Russell and Martin (2004), and Russell and Hall (2006), with kind permission). Compartmental pore space may have been partially filled with rapidly precipitated dendrites. The walls to the pores comprised nanocrystals of iron compounds, chiefly of FeS (Wolthers et al 2003) but including greigite, vivianite, and green rust occupying a silicate matrix. Tapping the ambient protonmotive force the pores and bubbles acted as catalytic culture chambers for organic synthesis, open to $H_2$, $NH_3$, $CH3^-$ at their base, selectively permeable and semi-conducting at their upper surface. The font size of the chemical symbols gives a qualitative indication of the concentration of the reactants.

**Figure 7 : Framboids in chimney**: Small pyrite vent structure: Reflected ore microscopy of transverse section shows a central area of empty black spaces plus (grey) fine framboidal pyrite, and a fine euhedral authigenic rim surrounded by baryte, with minor pyrite; (Picture by Dr. Adrian Boyce reproduced with his kind permission; Source: Boyce et al. 1983; Boyce, A.J. (1990). Exhalation, sedimentation and sulphur isotope geochemistry of the Silvermines Zn + Pb + Ba deposits, County Tipperary, Ireland: Unpublished *Ph.D. thesis*, Glasgow, U.K., University of Strathclyde, 354 p.).

**Supplementary Information: Figures and Legends**

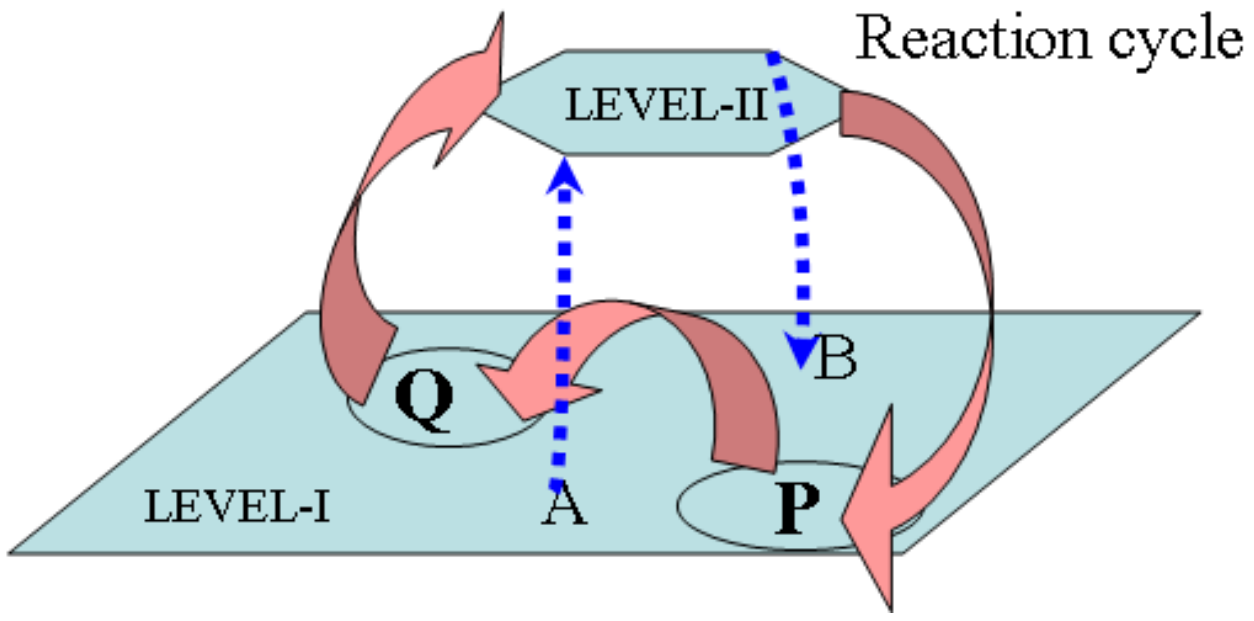

**Figure A**

The feedback-coupling between the control-network (level-I) and the metabolic-network (level-II), is extrapolated to the pre-biotic era to rephrase Orgel's (2000) concerns regarding plausible assumptions on the nature of minimal information- processing capabilities of mineral surfaces for hosting/organizing a proto-metabolic cycle. A capacity for interactions enabling long range energy and electron transfers (represented by bold orange and dashed blue arrows, respectively) is needed at level-I --the hosting surface, depicted as a green parallelogram,-- for proto-metabolic reaction cycles to organize at level-II—depicted as green hexagon. Did the host-surface at level-I have the ability to capture and channel the thermal energy released into it, say at point P (i.e. from an exothermic reaction taking place at level-II), to another spatio/temporal location, say point Q, where potential reactants (for endothermic reaction at level-II) could get recruited into the expanding cycle? Similarly electrons, required for some reactions of the cycle, would have led to exchanges (shown in blue dashed arrows) with the level-I catalytic colloids.

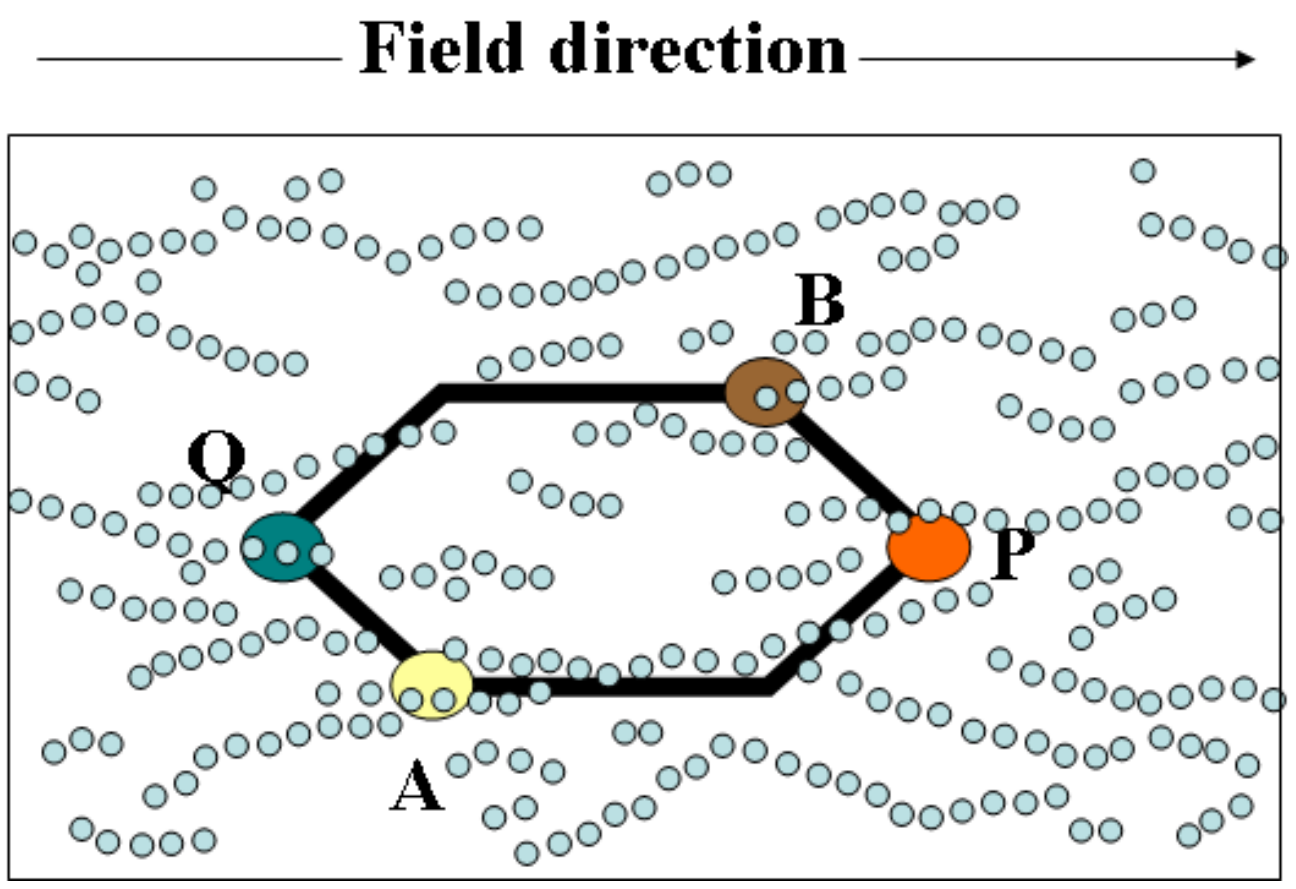

**Figure B**

The possibility of percolation through an inorganic network of dipolar interactions makes it interesting to consider a field-controlled network of magnetic mineral particles as a hosting surface to pre-biotic attractor cycles a la level-I. Figure B, is a top view of Figure A, where the green parallelogram representing the hosting surface is a “layer” of field-structured colloids, adapted from Figure 2, main text.

We speculate that transfers of electrons and heat energy through the dipolar network (Sect. 4.6) could drive the magnetic system out of equilibrium. This is since each individual particle’s composite magnetic moment in turn is directly affected by its redox state, and also the local temperature, thus affecting their collective dynamics. Taken above a threshold these feedback effects have the potential to cause phase-transitions to regimes with new types of collective ordering, leading to a long range correlation.